\magnification=\magstep1
\pageno=1                               
%
% Text - FINAL FINAL FINAL version after referee's comments, and Deb's
% typo changes on vinces request (24/2/99)
%
% the figures we have are:
%
% Fig 1: HH211 - I prefer the ``busy'' 3 panel one now. Gueth et al
% Fig 2: RNO43 CO 2-1 and Halpha
% Fig 3: G192.16
% Fig 4: m(v) plots
% Fig 5: energetics plots
% Fig 6: L1157 ISO H2 image (plus shocked stuff?)
\overfullrule0pt

\def\PCO{\hbox{$P_{\rm CO}$}}
\def\LCO{\hbox{$L_{\rm CO}$}}
\def\vCO{\hbox{$v_{\rm CO}$}}
\def\tdyn{\hbox{$\tau_{\rm d}$}}
\def\vw{\hbox{$v_{w}$}}
\def\vk{\hbox{$v_{K}$}}
\def\Mw{\hbox{$M_{w}$}}

\def\Fw{\hbox{$F_{w}$}}
\def\Mstar{\hbox{$M_{*}$}}
\def\Lstar{\hbox{$L_{*}$}}
\def\Rstar{\hbox{$R_{*}$}}
\def\FCO{\hbox{$F_{\rm CO}$}}
\def\MCO{\hbox{$M_{\rm CO}$}}
\def\Mwdot{\hbox{$\dot{M}_{w}$}}
\def\Madot{\hbox{$\dot{M}_{a}$}}
\def\kms{\hbox{km~s$^{-1}$}}
\def\Lsun{\hbox{L$_{\odot}$}}
\def\Msun{\hbox{M$_{\odot}$}}
\def\degree{\hbox{$^{\circ}$}}
\def\Halpha{\hbox{H$\alpha$}}
\def\Lbol{\hbox{$L_{\rm bol}$}}
\def\yr{\hbox{yr$^{-1}$}}
\def\mi{$\mu$m}
\def\cm3{cm$^{-3}$}
\def\percc{\hbox{${\rm cm}^{-3}$}}
\def\Htwo{\hbox{${\rm H}_2$}}
\def\micron{\hbox{$\mu{\rm m}$}}
% ppiv-style.tex
% Macro file for authors of chapter for PPIV
%
%-----------------------------------------------------------------------
% 
% FONT DEFINITIONS:
%
%-----------------------------------------------------------------------
% Basic font definitions borrowed from TeXbook macros
\font\ninerm=cmr9
\font\eightrm=cmr8
\font\sixrm=cmr6
\font\ninei=cmmi9
\font\eighti=cmmi8
\font\sixi=cmmi6
\skewchar\ninei='177 \skewchar\eighti='177 \skewchar\sixi='177
\font\ninesy=cmsy9
\font\eightsy=cmsy8
\font\sixsy=cmsy6
\skewchar\ninesy='60 \skewchar\eightsy='60 \skewchar\sixsy='60

\font\ninebf=cmbx9
\font\eightbf=cmbx8
\font\sixbf=cmbx6
\font\ninett=cmtt9
\font\eighttt=cmtt8
\hyphenchar\tentt=-1 % inhibit hyphenation in typewriter type
\hyphenchar\ninett=-1
\hyphenchar\eighttt=-1
\font\ninesl=cmsl9
\font\eightsl=cmsl8
\font\nineit=cmti9
\font\eightit=cmti8
\newskip\ttglue
\def\tenpoint{\def\rm{\fam0\tenrm}%
  \textfont0=\tenrm \scriptfont0=\sevenrm \scriptscriptfont0=\fiverm
  \textfont1=\teni \scriptfont1=\seveni \scriptscriptfont1=\fivei
  \textfont2=\tensy \scriptfont2=\sevensy \scriptscriptfont2=\fivesy
  \textfont3=\tenex \scriptfont3=\tenex \scriptscriptfont3=\tenex
  \def\it{\fam\itfam\tenit}%
  \textfont\itfam=\tenit
  \def\sl{\fam\slfam\tensl}%
  \textfont\slfam=\tensl
  \def\bf{\fam\bffam\tenbf}%
  \textfont\bffam=\tenbf \scriptfont\bffam=\sevenbf
   \scriptscriptfont\bffam=\fivebf
  \def\tt{\fam\ttfam\tentt}%
  \textfont\ttfam=\tentt
  \tt \ttglue=.5em plus.25em minus.15em
  \normalbaselineskip=12pt
  \let\sc=\eightrm
  \let\big=\tenbig
  \setbox\strutbox=\hbox{\vrule height8.5pt depth3.5pt width0pt}%
  \normalbaselines\rm}
\def\ninepoint{\def\rm{\fam0\ninerm}%
  \textfont0=\ninerm \scriptfont0=\sixrm \scriptscriptfont0=\fiverm
  \textfont1=\ninei \scriptfont1=\sixi \scriptscriptfont1=\fivei
  \textfont2=\ninesy \scriptfont2=\sixsy \scriptscriptfont2=\fivesy
  \textfont3=\tenex \scriptfont3=\tenex \scriptscriptfont3=\tenex
  \def\it{\fam\itfam\nineit}%
  \textfont\itfam=\nineit
  \def\sl{\fam\slfam\ninesl}%
  \textfont\slfam=\ninesl
  \def\bf{\fam\bffam\ninebf}%
  \textfont\bffam=\ninebf \scriptfont\bffam=\sixbf
   \scriptscriptfont\bffam=\fivebf
  \def\tt{\fam\ttfam\ninett}%
  \textfont\ttfam=\ninett
  \tt \ttglue=.5em plus.25em minus.15em
  \normalbaselineskip=10pt % set to 10pt, not standard 11pt of TeX manmacs
  \let\sc=\sevenrm
  \let\big=\ninebig
  \setbox\strutbox=\hbox{\vrule height8pt depth3pt width0pt}%
  \normalbaselines\rm}
\def\eightpoint{\def\rm{\fam0\eightrm}%
  \textfont0=\eightrm \scriptfont0=\sixrm \scriptscriptfont0=\fiverm
  \textfont1=\eighti \scriptfont1=\sixi \scriptscriptfont1=\fivei
  \textfont2=\eightsy \scriptfont2=\sixsy \scriptscriptfont2=\fivesy
  \textfont3=\tenex \scriptfont3=\tenex \scriptscriptfont3=\tenex
  \def\it{\fam\itfam\eightit}%
  \textfont\itfam=\eightit
  \def\sl{\fam\slfam\eightsl}%
  \textfont\slfam=\eightsl
  \def\bf{\fam\bffam\eightbf}%
  \textfont\bffam=\eightbf \scriptfont\bffam=\sixbf
   \scriptscriptfont\bffam=\fivebf
  \def\tt{\fam\ttfam\eighttt}%
  \textfont\ttfam=\eighttt
  \tt \ttglue=.5em plus.25em minus.15em
  \normalbaselineskip=9pt
  \let\sc=\sixrm
  \let\big=\eightbig
  \setbox\strutbox=\hbox{\vrule height7pt depth2pt width0pt}%
  \normalbaselines\rm}
%
% Now in a position to define font sizes for book headers, captions, etc.
\def\headtype{\ninepoint}                 % headers
\def\abstracttype{\ninepoint}             % abstracts
\def\captiontype{\ninepoint}              % figure captions
            % table notes
\def\footnotetype{\ninepoint}             % footnotes
                  % references
\def\refit{\it}                           % italics in references
\font\chaptitle=cmr10 at 11pt             % chapter title font
\rm                                       % make sure we load cmr fonts

%-----------------------------------------------------------------------
%
% GENERAL SPACINGS AND SKIPS:
%
%-----------------------------------------------------------------------
% spacings and spacing definition
\parindent=0.25in                         % paragraph indentation
\parskip=0pt                              % extra skip between paragraphs
\baselineskip=12pt                        % skip between lines
\hsize=4.25truein                         % width of text
\vsize=7.445truein                        % height of text; exactly 45 lines 
\hoffset=1in                              % horizontal offset on page
\voffset=-0.5in                           % vertical offset on page

% define skips before and after sections etc.
\newskip\sectionskipamount                % skip before main section
\newskip\aftermainskipamount              %      after  main section
\newskip\subsecskipamount                 %      before subsection
\newskip\firstpageskipamount              %      at top of first page
\newskip\capskipamount                    %      in captions
\newskip\ackskipamount                    %      before acknowledgments
\sectionskipamount=0.2in plus 0.09in
\aftermainskipamount=6pt plus 6pt         % needs to be as much as whole line
\subsecskipamount=0.1in plus 0.04in
\firstpageskipamount=3pc
\capskipamount=0.1in
\ackskipamount=0.15in
\def\sectionskip{\vskip\sectionskipamount}
\def\aftermainskip{\vskip\aftermainskipamount}
\def\subsecskip{\vskip\subsecskipamount} 

\def\capskip{\hskip\capskipamount}

%-----------------------------------------------------------------------
%
% PAGINATION AND HEADING MACROS:
%
%-----------------------------------------------------------------------
% pagination and related macros
\nopagenumbers                            % turn off default TeX numbering
\newcount\firstpageno                     % create count to hold first page no.
\firstpageno=\pageno                      % remember first page no.: pageno
                                          % should have been set as first line
                                          % in chapter before macros are read
\newcount\chapno                          % create count to hold chapter no.

% all pages have a running head, its contents depending on whether
% page is right or left facing. Exceptions are the first page of
% each chapter and pageinserts (not available to authors, only
% for SSSBOOKS use): these have no head
\def\rightheadline{\headtype\phantom{\folio}\hfil\runningtitletext\hfil\folio}
\def\leftheadline{\headtype\folio\hfil\runningauthortext\hfil\phantom{\folio}}
\headline={\ifnum\pageno=\firstpageno\hfil
           \else
              \ifdim\ht\topins=\vsize           % a full pageinsert
                 \ifdim\dp\topins=1sp \hfil     % a rotated full pageinsert
                 \else
                     \ifodd\pageno\rightheadline\else\leftheadline\fi
                 \fi
              \else
                 \ifodd\pageno\rightheadline\else\leftheadline\fi
              \fi
           \fi}

% a footline page number is used for the first page of a chapter only
\def\bottomnumber{\hss\tenrm[\folio]\hss}
\footline={\ifnum\pageno=\firstpageno\bottomnumber\else\hfil\fi}

%-----------------------------------------------------------------------
%
% FORMATS FOR SECTIONS, ABSTRACTS, CAPTIONS, FOOTNOTES, ETC.:
%
%-----------------------------------------------------------------------
% macros to create sections and subsections - note that contrary to good
% TeX practice, the macro parameters are _deliberately_ delimited by
% whitespace, and as a consequence, there must be nothing attached to
% the end of section argument in curly braces. The reason I did it this
% way are obscure, and probably not `good'. It looks like a kludge; it
% probably is. So be it. It works, as long as you follow the rules,
% i.e. use 
%
%  %
%  \mainsection{This is a main section heading}
%  %
%  Here is the text ...
%
\outer\def\mainsection#1
    {\vskip 0pt plus\smallskipamount\sectionskip
     \message{#1}\vbox{\noindent{\bf#1}}\nobreak\aftermainskip\noindent}
 
\outer\def\subsection#1
    {\vskip 0pt plus\smallskipamount\subsecskip
     \message{#1}\vbox{\noindent{\bf#1}}\nobreak\smallskip\nobreak\noindent}
 
% use this when subsection immediately follows mainsection heading
\def\backup{\nobreak\vskip-\baselineskip\nobreak\vskip-\subsecskipamount\nobreak
}

% macros for formatting chapter title, author names, affiliation, abstract, 
% references, figure captions 
\def\title#1{{\chaptitle\leftline{#1}}}
\def\name#1{\leftline{#1}}
\def\affiliation#1{\leftline{\it #1}}
\def\abstract#1{{\abstracttype \noindent #1 \smallskip\vskip .1in}}
\def\ref{\noindent \parshape2 0truein 4.25truein 0.25truein 4truein}
\def\caption{\noindent \captiontype
             \parshape=2 0truein 4.25truein .125truein 4.125truein}

% version of footnote stuff copied from TeXbook format (as opposed to 
% plain TeX version)
\def\footnote#1{\edef\fspafac{\spacefactor\the\spacefactor}#1\fspafac
      \insert\footins\bgroup\footnotetype
      \interlinepenalty100 \let\par=\endgraf
        \leftskip=0pt \rightskip=0pt
        \splittopskip=10pt plus 1pt minus 1pt \floatingpenalty=20000
        \textindent{#1}\bgroup\strut\aftergroup\strut\egroup\let\next}
\skip\footins=12pt plus 2pt minus 4pt % space added when footnote is present
\dimen\footins=30pc % maximum footnotes per page

%-----------------------------------------------------------------------
%
% OTHER GENERAL MACROS:
%
%-----------------------------------------------------------------------
% shorthand for \noindent

% macro to force end of sentence space after capital letters with period
\def\@{\spacefactor 1000}

% redefine \, to work in both math and lr mode
\def\,{\pcomma} 
\def\pcomma{\relax\ifmmode\mskip\thinmuskip\else\thinspace\fi}

% define a reasonable version of \simgt and \simlt using \mathpalette
% as explained in the TeXbook p151 and in the definition of \cong in
% Appendix B. Only problem with these is that the vertical spacing
% commands defined in terms of ex do not seem to reflect the actual
% font sizes in \scriptstyle etc. as expected
\def\oversim#1#2{\lower0.5ex\vbox{\baselineskip=0pt\lineskip=0.2ex
     \ialign{$\mathsurround=0pt #1\hfil##\hfil$\crcr#2\crcr\sim\crcr}}}

\input psfig
\parindent=0.25in
\parskip=0pt     
\baselineskip=12pt
\hsize=6.0truein  
\vsize=9.0truein  
\hoffset=0.125in 
\voffset=-0.0in
\long\def\caption#1{\vskip6pt{\noindent\it #1}\par\vskip1pc}
\def\ref{\noindent \parshape2 0truein 6.0truein 0.25truein 5.75truein}

\def\runningtitletext{MOLECULAR OUTFLOWS}
\def\runningauthortext{Richer, Shepherd, Cabrit, Bachiller and Churchwell}

\null
%\firstpageskip
{\it \noindent To appear in Protostars \& Planets IV, eds. V. Mannings, A. Boss,
S. Russell (Tucson: The University of Arizona Press)}\vskip6mm

{\baselineskip=14pt
\title{MOLECULAR OUTFLOWS FROM YOUNG STELLAR OBJECTS}
}

\vskip .3truein
\name{JOHN RICHER}
\affiliation{Cavendish Laboratory, Cambridge, UK}
\vskip .2truein
\name{DEBRA SHEPHERD}
\affiliation{California Institute of Technology}
\vskip .2truein
\name{SYLVIE CABRIT}
\affiliation{DEMIRM, Observatoire de Paris}
\vskip .2truein
\name{RAFAEL BACHILLER}
\affiliation{IGN Observatorio Astron\'omico Nacional}
\vskip .1truein
\leftline{and}
\vskip .1truein
\name{ED CHURCHWELL}
\affiliation{University of Wisconsin, Madison}
\vskip .3truein

\abstract{
We review some aspects of the bipolar molecular outflow phenomenon. In
particular, we compare the morphological properties, energetics and
velocity structures of outflows from high and low-mass protostars and
investigate to what extent a common source model can explain outflows
from sources of very different luminosities.  Many flow properties, in
particular the CO spatial and velocity structure, are broadly similar
across the entire luminosity range, although the evidence for
jet-entrainment is still less clear cut in massive flows than in
low-mass systems.  We use the correlation of flow momentum deposition
rate with source luminosity to estimate the ratio $f$ of mass ejection
to mass accretion rate. From this analysis, it appears that a common
driving mechanism could operate across the entire luminosity
range. However, we stress that for the high-mass YSOs, the detailed
physics of this mechanism and how the ejected wind/jet entrains
ambient material remain to be addressed. We also briefly consider the
alternative possibility that high-mass outflows can be explained by
the recently proposed circulation models, and discuss several
shortcomings of those models. Finally, we survey the current evidence
on the nature of the shocks driven by YSOs during their
pre-main-sequence evolution.
}

\mainsection{I.~~INTRODUCTION}
\backup
Stars of all masses undergo energetic, generally bipolar, mass
loss during their formation.  While diagnostics of mass loss using
optical and near-infrared line emission provide the most dramatic
images of the violent birth of stars, molecular flows,
predominantly traced by their CO emission lines, provide the strongest
constraints on models of cloud collapse and star formation. This is
because molecular flows are the massive and predominantly cool
reservoir where most of the flow momentum is eventually deposited and
so provide a fossil record of the mass loss history of the protostar.
In contrast, the optical and near-infrared emission arises from hot
shocked gas which cools in a few years, and hence only traces the
currently active shocks in the flow.  In addition, since there is
negligible extinction at millimeter wavelengths, even the youngest,
deeply embedded objects can be studied.

Since the last review of this subject in the previous conference
proceedings in this series (Fukui et al.\ 1993), many breakthroughs in
understanding the nature of these objects have been forthcoming,
primarily as a result of improvements at millimeter and submillimeter
wavelength observatories. (We refer the reader in particular to  
volume 1 of the conference series Revista Mexicana de Astronom{\'\i}a 
y Astrof{\'\i}sica and the proceedings of the IAU Symposium 182).
More sensitive receivers and focal
plane arrays have allowed wide-field imaging of outflows, showing that
they often extend to many parsecs in size, sometimes even beyond their
natal cloud boundaries (e.g. Padman et al.\ 1997).  In addition, the
availability of mosaiced interferometric images of many outflows at
$1-2''$ resolution has led to significant breakthroughs in
understanding the small-scale structure of outflows (e.g. Bachiller et
al.\ 1995; Cernicharo and Reipurth 1996; Gueth et al.\ 1996) and
provide strong constraints on viable outflow-acceleration models, in
particular on the role of jets and the nature of the entrainment
mechanism (Cabrit et al.\ 1997; Shu and Shang 1997).  Finally,
multi-line studies (Bachiller and P\'erez Guti\'errez 1997) have shown
that shocks driven by flows chemically process the ISM at a rapid
rate, modifying its composition and having profound implications for
chemical and dynamical models of molecular clouds.

In this chapter, we review some of the recent results on outflows from
both low and high-mass protostars and focus on a comparison of their
physical properties. In section II, we summarize the observed flow
morphologies and velocity structures, and look for evidence that a
common mechanism could reproduce the observations of all such flows.
In section III, the possible entrainment and ejection mechanisms for
outflows are addressed, and the evidence for a common mechanism is
discussed based on an analysis of the flow energetics.  Finally, in
section IV, the evidence for, and nature of, shocks in outflows is
presented.

\mainsection{{I}{I}.~~OUTFLOW STRUCTURE}

Molecular outflows come in a wide variety of shapes and sizes. This is
hardly surprising given the likely diversity in the mass and
multiplicities of the driving protostars, the outflow ages, and
molecular environments. Although at least two hundred outflows have
been cataloged (Wu et al.\ 1996), there is still no large homogeneous
sample which has been mapped with good resolution and sensitivity, so
much of our current understanding is heavily influenced by a few
well-studied examples. This is even more true for high-mass systems,
where perhaps only 10 or so outflows have been studied in detail,
although rapid progress is being made using millimeter-wave
interferometers to study flows in the often complex environment of
high-mass star formation (Shepherd et al.\ 1998).  Consequently, our
current estimates of ``typical'' outflow properties may be far from
accurate.

\subsection{A.~~Low-Mass Systems}

It is now clear that stellar jets, with speeds of 100-300\,\kms\ and
densities of order $10^3\,\percc$, are responsible for accelerating
much of the molecular gas in many of the youngest low-mass outflow
systems (e.g. Richer et al.\ 1992; Padman and Richer 1994; Bachiller 
et al.\ 1995). However, there is also good evidence for momentum being
deposited into the flows by wind components with wider opening angles,
and that this component perhaps becomes relatively more powerful as
the flows age (e.g. Bence et al.\ 1998).

$$\psfig{file=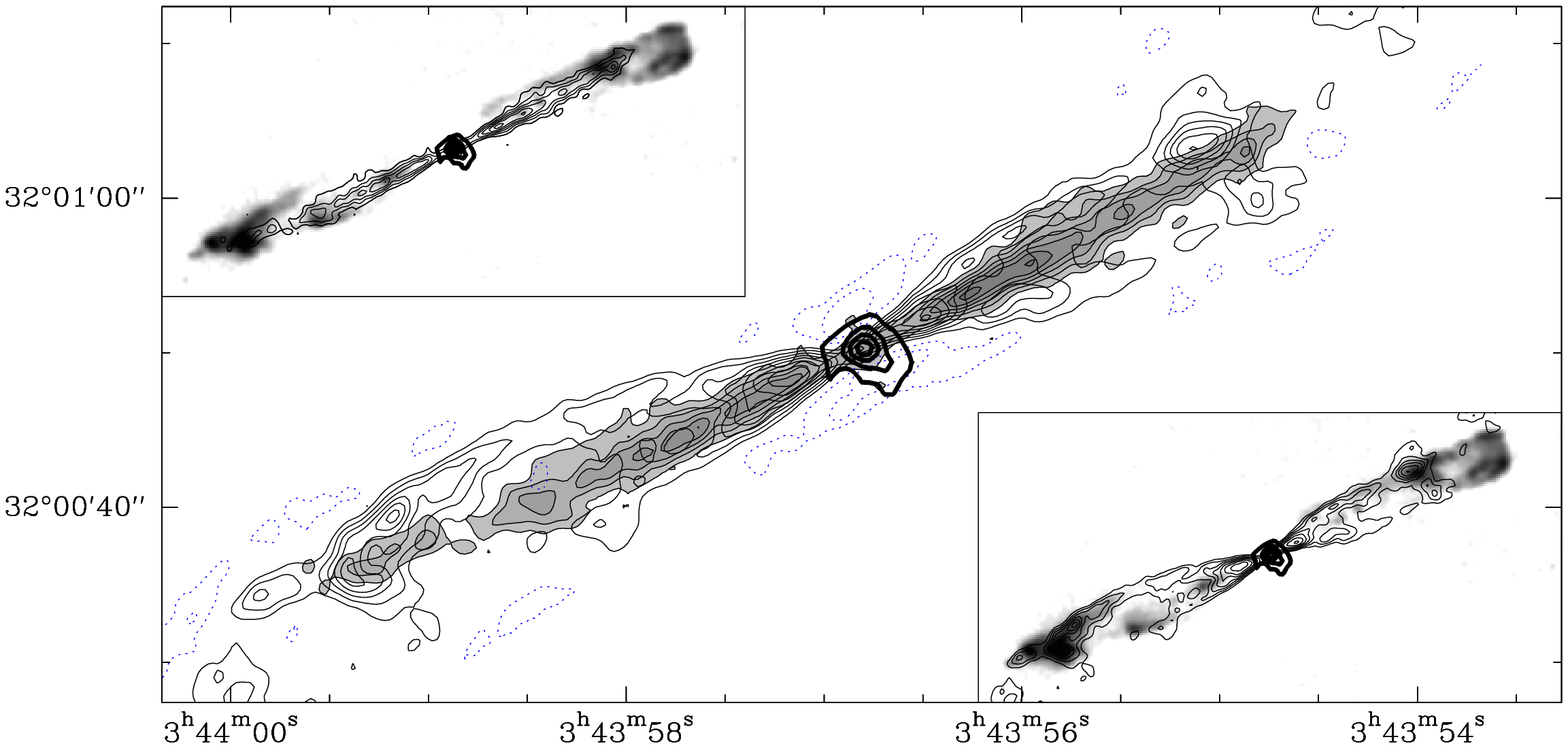,angle=270,width=5.2in}$$
\caption{Figure~1.\capskip The HH211 molecular jet mapped with the
Plateau de Bure interferometer. The main panel shows the high-velocity
CO jet in greyscale superposed on the lower velocity outflowing gas,
which forms a cavity around the jet. The thick contours show the 1.3mm
continuum emission. The panels top left and lower right show the fast
and slow CO emission overlaid on the shocked H$_2$ line emission. Data
taken from McCaughrean and Zinnecker 1996, and Gueth and Guilloteau 1999.}

Some of the first clear evidence for a jet-dominated outflow origin
was identified in the flows from L1448C (Bachiller et al.\ 1990;
Guilloteau et al.\ 1992; Dutrey et al.\ 1997) and NGC2024-FIR5 (Richer
et al.\ 1989, 1992). In L1448, a spectacular SiO jet is seen emanating
from the driving source, aligned with the large-scale CO flow and
unresolved across its width.  In the NGC2024-FIR5 flow, the fastest
gas is seen to form an elongated jet-like feature on the axis of
nested shells of lower velocity gas. The collimation ratio $q$,
defined as the width of the flow to the distance to the driving
source, is as high as 30 for the high velocity (30\,\kms) gas in this
source, but only 4 or so for the lower velocity (5\,\kms) envelope.
Many other flows are now known which show this structure, with
high-velocity elongated components tracing jet-like activity lying
inside cavities of lower velocity gas; examples include L1157 (Gueth
et al.\ 1996), HH111 (Cernicharo and Reipurth 1996; Nagar et al.\
1997) and HH211 (Gueth and Guilloteau 1999). 
All of these flows appear to be
driven by very young, low-mass objects, based on their low
luminosities, and non-detection at even infrared wavelengths. The CO
flow from HH211 (Fig.\ 1) is perhaps the most striking image to date
of this phenomenon, showing an unresolved CO jet with high velocity
gas ($v>10\,$\kms\ with no correction for inclination) lying within an
ovoid cavity of slower moving gas ($<10\,$\kms).  HH211 is probably
also one of the youngest low-mass outflows known, having a {\it
dynamical age} $\tdyn = 0.07{\rm pc} / 10\,\kms = 7000$ years; if the
source is inclined close to the plane of the sky, as is expected
given the clear separation of red and blue outflow lobes and the
relatively modest projected flow speeds, the true dynamical age could
well be a factor of 5 or so lower. It appears that HH211 is 
at the very start of its main accretion phase, and this perhaps
explains the relative simplicity of the outflow structure. Note that
the full opening angle at the base of the flow is only 22\degree.

It is natural to identify these so-called ``molecular jets'' as the
deeply-embedded counterparts of the Herbig-Haro (HH) jets seen in less
obscured systems and as the neutral counterparts of the ionized jets
seen in the radio (see chapters by Eisl\"offel et al., and 
by Hartigan et al., this volume).   Important work by Raga (1991) and 
Hartigan et al.\ (1994) demonstrated that HH jets were
denser and hence more powerful than initial estimates suggested
(e.g. Mundt et al.\ 1987), so that the total momentum flux in HH jets
integrated over the lifetime of a typical source is sufficient to
drive the observed CO flows (e.g. Richer et al.\ 1992; Mitchell et
al.\ 1994).  This unified picture of jet-driven outflows is entirely
consistent with the observed CO structures discussed above.  (Masson
and Chernin 1993; Cabrit et al. 1997; Smith et al. 1997; Gueth and
Guilloteau 1999).  However, we caution that the actual composition of
the driving jet, whether primarily atomic or molecular, is unknown.
It is still unclear whether the CO jets seen in sources such as HH211
and NGC2024-FIR5 arises (1) from the body of a jet where molecules
have formed in the gas phase (Glassgold et al.\ 1989; Smith et al.\
1997); (2) from molecules formed in the post-shock region of shocks in
the jet; or (3) from ambient molecular gas turbulently entrained along
the jet's edge and at internal working surfaces (Raga et al.\ 1993;
Taylor and Raga 1995).

$$\psfig{file=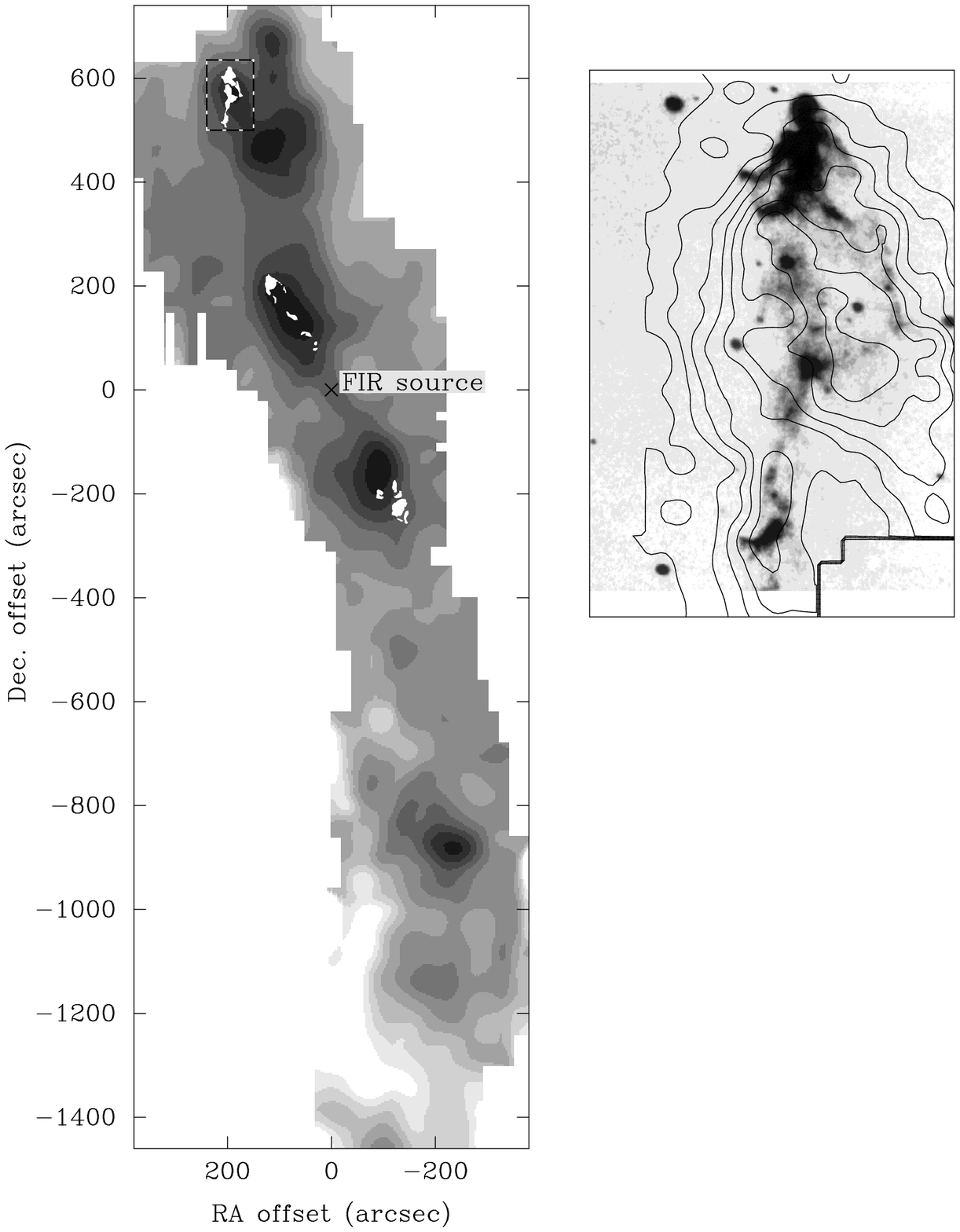,angle=0,height=120mm}$$
\caption{Figure~2.\capskip CO 2-1 image (in greyscale) of the RNO~43 molecular
outflow (Bence at al.\ 1996); the driving FIR source is at the origin,
marked with a cross. The solid white patches show the extent of the
\Halpha\ emission. Note the large flow extent and correlation of the
\Halpha\ emission with the CO hotspots. The panel to the right shows a
detail from the northernmost \Halpha\ emission patch, showing the
bow-shock shaped \Halpha\ structure in greyscale, and CO 2-1 contours
overlaid. }

RNO43-FIR is another low-luminosity outflow source, and is probably a flow
in middle-age (Bence et al.\ 1996; Cabrit et al.\ 1988): the driving
source, with $\Lstar=6$ \Lsun , is a heavily embedded Class~0 protostar
invisible even at 2\micron, but the flow extends over a
total size of 3\,pc. The outflow axis is close to the plane of the sky, so
the image of CO integrated intensity shown in Fig.\ 2 shows both blue and
red-shifted sides of the flow. The dynamical age is $\tdyn \sim 1.5\,{\rm
pc} / 10\,\kms = 1.5\times10^5$ years which is an upper limit due to the
small but unknown inclination angle. The CO flow is much more complex than
HH211 and this most likely reflects the clumpy nature of the molecular gas
through which the driving jet has passed; nonetheless, a very approximate
S-shape symmetry can be seen about the driving source, both in the spatial
and velocity data, suggesting that the jet direction has changed over time
(Bence et al.\ 1996).  A precessing jet model, 
with a full cone angle of
27\degree\ and a precession timescale of order $3\times10^4$ years, provides
a reasonable description of the overall flow properties (Bence et al.\
1996); such precession could be driven by a
binary companion in an inclined orbit. 

The evidence that this flow is jet-driven even at a distance of
1.5\,pc from the driving source is demonstrated by the \Halpha\ images
of the object, which show strong emission coincident with the
brightest CO points (see Fig.\ 2). In addition, at the northernmost CO
feature in the flow, bright
\Halpha\ coincident with a bow-shock shaped CO feature
suggests that the whole of the RNO43 outflow can be explained by the
gradual sweeping up of a clumpy molecular cloud by a powerful stellar
jet whose ejection axis varies slowly with time. It is also
interesting to note that parsec-scale flows such as RNO43 are the
natural counterparts to the parsec-scale Herbig-Haro flows (Reipurth
et al.\ 1997) seen in wide-field optical imaging: if there is
molecular gas in the jet's path, then flows such as RNO43 result,
whereas jets in essentially empty space such as HH34 (Devine et al.\
1997) show only optical emission.  This points strongly to the jets
being primarily atomic in composition.  However, in the HH111 outflow
system, CO ``bullets'' associated with the optical jet and having similar
velocities to the optical gas have been detected far beyond the
molecular cloud boundary (Cernicharo and Reipurth 1996). This suggests
that in some cases the jets may have a molecular component.

Not all low-mass flows are jet-dominated: many of the older flows show
CO emission dominated by low-velocity cavities with little evidence
for elongated high-velocity CO jet features. Examples include L43
(Bence et al.\ 1998), L1551 (Moriarty-Schieven et al.\ 1988) and B5
(Velusamy and Langer 1998). These flows are typically
$10^{5-6}\,$years old, and associated with nebulosities visible in the
optical or at 2\,\micron, suggesting they are Class~I or II
protostars, older than the Class~0 objects responsible for the HH211,
RNO43 and NGC2024-FIR5 outflows. 
The lack of high-velocity CO and
obvious jet-like features in these flows, in contrast to their younger
counterparts such as HH211, and the presence of much wider cavities at
the base of the flows, suggests that the jet power has declined over
time, and that a wider opening angle wind is now primarily responsible
for driving the outflow.  However, even in these older sources there
is usually some evidence for weak jet activity: L1551 has an optical
jet and extended Herbig-Haro emission apparently on one of its cavity
walls (Davis et al.\ 1995; Fridlund and Liseau 1998), as well as fast
CO emission suggestive of a jet-origin (Bachiller et al.\ 1994), and
B5 has optical jets stretching 2.2 pc away from the driving
source. However, the L43 flow shows no signs of shocks, jets or very
fast CO, and this may be a true ``coasting'' flow, which is no longer
being accelerated by a stellar wind or jet.

The recent interferometric images of the B5 outflow 
(Velusamy and Langer 1998)
reveal a beautiful example of wide, hollow cavities at the base of
the outflow, much like the reflection nebulosities seen in the 
near-infrared in many of these systems; in order to reproduce the clearly
separated red- and blue-shifted emission lobes, the CO must be flowing
along these cavity walls, presumably being accelerated by a
poorly-collimated radial wind from the star.  The entire outflow
driven by B5 has a dynamical age of $10^6\,$years, and a collimation
factor of about 5.  The cavity opening angle at the protostar is
extremely large, in the range $90-125\degree$, and Velusamy and Langer
(1998) suggest that if this angle further broadens with time it may
ultimately cut off the accretion flow. This idea is consistent with
most of the available data: the young flows such as HH211, RNO43 and
L1157 have opening angles less than 30\degree or so, while the older
flows such as L1551, L43 and B5 are significantly greater than
90\degree. With maps at good resolution of a larger sample of
outflows, it will be possible to test the hypothesis that flow opening
angle is a measure of the source age.

\subsection{B.~~High-Mass Systems}

Our understanding of massive flows is beginning to change because we
are starting to find a few isolated systems that can be studied in
depth.  However, we are still observationally biased toward older
flows that are easier to identify and study.  This bias is likely to
diminish in the future as more massive young flows are identified
(e.g. Cesaroni et al.\ 1997; Molinari et al.\ 1998;  Zhang et al.\ 1998).

Most luminous YSOs have relatively wide-opening angle outflows as
defined by their CO morphology.  Although the statistics are poor
because few massive flows have been studied with sufficient resolution
to adequately determine the morphology, collimation factors {\it q}
for 7 well-mapped flows produced by YSOs with $\Lbol > 10^3\,\Lsun$
range from 1 to 1.8 (NGC 7538 IRS1: Kameya et al.\ 1989; 
HH~80-81: Yamashita et al.\ 1989; NGC 7538 IRS9: Mitchell et al.\ 1991; 
GL 490: Mitchell et al.\ 1995; Ori~A : Chernin and Wright 1996; 
W75N: Davis et al.\ 1998; G192.16: Shepherd et al.\ 1998).  
The dynamical time
scales for these outflows range from 750 years to $\sim 2 \times 10^5$
years and there is no obvious dependence of flow collimation on
age. In comparison, collimation factors in low-mass outflows range
from $\sim 1$ to as high as 10 with a typical value being $\sim$~2 or
3 (Fukui et al.\ 1993 and references therein).  It appears that more
luminous YSOs do not in general produce very well-collimated CO 
outflows and this result is independent of outflow age, unlike
outflows produced by low-luminosity YSOs.  This may be due to the fact
that outflows from luminous YSOs tend to break free of their molecular
cloud core at a very early stage in the outflow process.  Hence,
massive molecular flows are frequently the truncated base of a much
larger outflow that extends well beyond the cloud boundaries
(e.g.  HH~80--81: Yamashita et al.\ 1989; 
DR21: Russell et al.\ 1992; G192.16: Devine et al.\ 1999).

The outflow from Orion~A is perhaps the best known and best
studied high-mass outflow system.  The driving source is believed to
be an O star, and the estimated age of the flow is $\sim 750$ years
which makes this one of the youngest known outflows.  The flow differs
significantly from those produced by low-luminosity sources and
represents a spectacular example of the outflow phenomenon.  The CO
outflow is poorly collimated at all velocities and spatial resolutions
and the H$_2$ emission is dispersed into a broad fan shape that is
unlike any other known outflow. Its morphology is highly suggestive of
an almost isotropic, explosive origin.  McCaughrean and Mac~Low (1997)
model the H$_2$ bullets as a fragmented stellar wind bubble using the
fragmentation model of Stone et al.\ (1995) and suggest that the bullets
are caused by several young sources within the BN-KL cluster.  Chernin
and Wright (1996) argue that the flow is driven by a single massive
YSO, source I.  The estimated opening angle, corrected for
inclination, is approximately 60\degree\ for the blue lobe and 120\degree\
for the red lobe.

$$\psfig{file=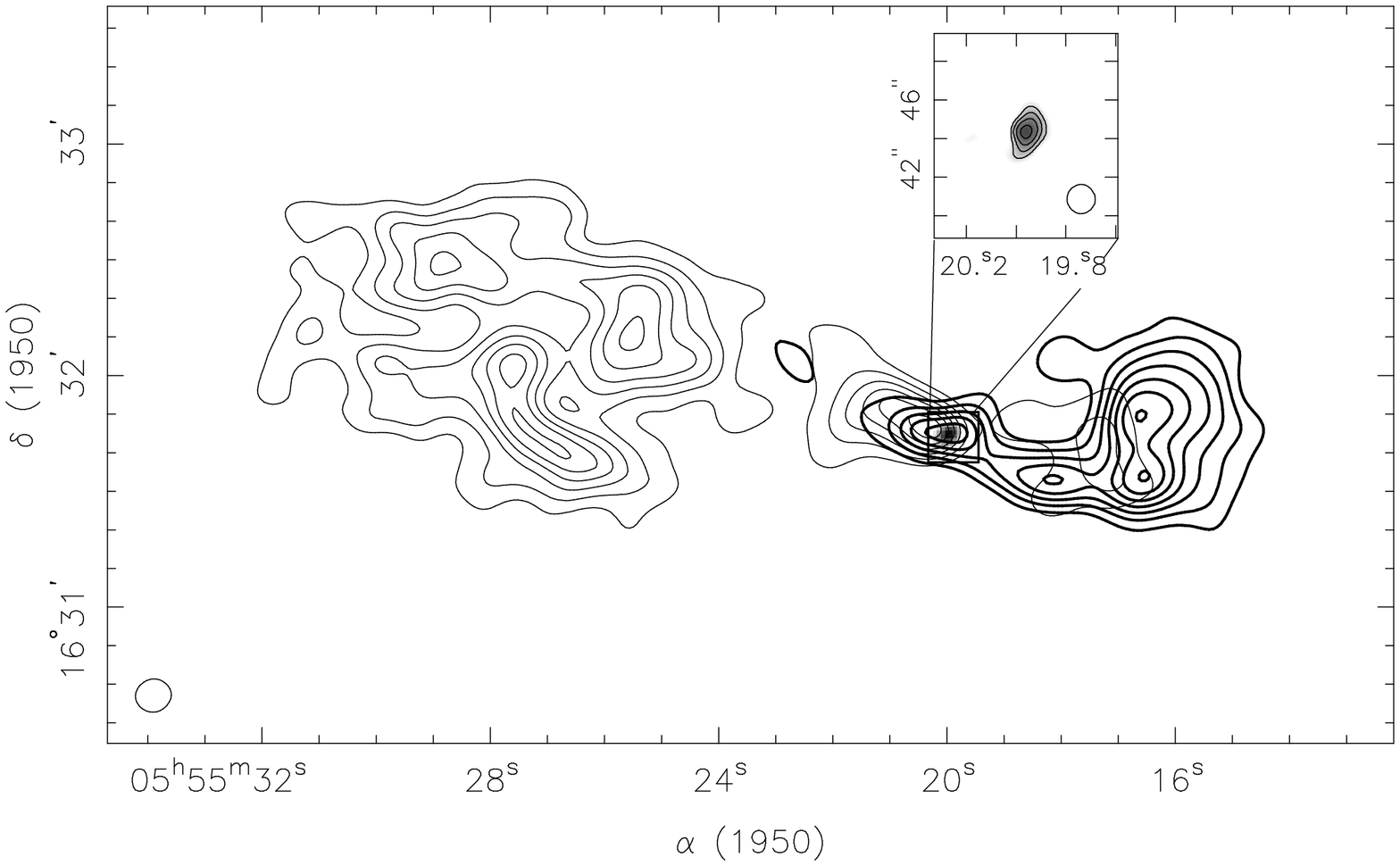,angle=270,width=4.8in}$$
\caption{Figure~3.\capskip The G192.16 outflow mapped with the OVRO
interferometer. Contours of redshifted CO (thick lines) and blueshifted
CO (thin lines) delineate the bipolar flow emanating from a dense core traced
by the central dust continuum emission represented in greyscale. }

The outflow from G192.16 is perhaps more typical of outflows from B
stars.  Figure 3 shows an interferometric CO J=1-0 image of the
outflow together with a closeup of the 3~mm continuum emission showing
a flattened distribution of hot dust (Shepherd et al.\ 1998; Shepherd
and Kurtz 1999).  The 3~pc long molecular flow represents the
truncated base of a much larger flow identified by \Halpha\ and [SII]
emission that extends almost 5~pc from the YSO (Devine et al.\ 1998).
The CO flow is $\sim 10^5$ years old and appears to be driven by an
early B star.  Despite the very different masses and energies
involved, the CO morphology looks very similar to L1551-IRS5: the
extent of the G192.16 CO outflow is $2.6 \times 0.8$~ pc,
approximately twice that of L1551, but the mass in the outflow
($\approx95\,\Msun$) is much greater.  The opening angle at the base of the
flow is $\sim 90\degree$.

The present lack of well-collimated CO outflows with $q >$ 3 from YSOs
with $\Lbol > 10^3\Lsun$ does not mean that jets and
well-collimated structures are not present in these massive sources.
For example, the central source in HH~80-81 ($\Lbol\sim 2
\times 10^4$ L$_{\odot}$) powers the largest known Herbig-Haro jet
with a total projected length of 5.3 pc, assuming a distance of 1.7 kpc
(Mart{\'\i} et al.\ 1993, 1995 and references therein).  
However, the CO flow appears poorly collimated with $q\sim 1$ when mapped
at moderate resolution (Yamashita et al.\ 1989).
Also, the biconical thermal radio jet from Cepheus A HW2 
($\Lbol \sim 10^4$ L$_{\odot}$) 
appears to be responsible for at least part of the
complicated molecular flow seen in CO and shock-enhanced species such as
H$_2$, SiO, and SO (e.g., Doyon and Nadeau 1988, Mart{\'\i}n-Pintado et
al. 1992, Hughes 1993, Torrelles et al. 1993,  Rodr{\'\i}guez et al. 1994,
Rodr{\'\i}guez 1995, Garay et al. 1996, Hartigan et al. 1996, Narayanan and
Walker 1996). 
The HH~80-81 and Cepheus A HW2 systems demonstrate that
high-luminosity YSOs can produce well-collimated jets like those found
in association with less luminous stars, even though the CO
flow may appear chaotic or poorly collimated. Other examples of possible jets
in massive outflows include IRAS~20126 and W75N IRS1.  IRAS~20126 ($\Lbol
\sim 1.3 \times 10^4\,\Lsun$) appears to drive a compact jet seen in SiO
and H$_2$ (Cesaroni et al.\ 1997). W75N IRS1 ($\Lbol \sim 1.4 \times
10^5\,\Lsun$) shows H$_2$ 2.12\mi\ shock-excited emission at the end
and sides of the CO lobe, with a morphology and emission
characteristics highly suggestive of a jet bowshock (Davis et al.\
1998). However, this ``bowshock'' is 0.3\,pc wide, i.e. 30 times
larger than the H$_2$ 2.12\mi\ bowshock at the end of the HH211 flow
(cf. Fig.~1). It is not fully clear whether such wide bows are created
by protostellar jets, or by a low collimation wind component. Scaling
up from current hydrodynamical jet simulations (e.g. Suttner et
al. 1997), one would need a jet radius of $\sim$ 0.03\,pc at 1.3\,pc
from the star, hence a jet opening angle $\sim$ 2.6$^{\circ}$.

\subsection{C.~~Outflow Velocity Structure}

From the above discussion, we conclude that jet activity, bow shocks, and
CO cavities are common to outflows from low and high-mass systems.  There
is some evidence that high-mass systems are less well
collimated than low-mass ones, but this may be due to selection effects ---
young, high-mass systems with small opening angles may simply be missing
from the small sample currently known (although the Orion outflow does
appear to be very young). This conclusion suggests it is sensible to
consider the possibility that a common driving mechanism is responsible for
all outflows.

Several authors have noted that molecular flows seem to be
characterized by a power-law dependence of flow mass $\MCO (v)$ as a
function of velocity. The power-law exponent $\gamma$ (where $\MCO (v)
\propto v^{\gamma}$) is typically $\sim -1.8$ for most low-mass
outflows, although the slope often steepens at velocities greater than
10~{\kms} from $v_{LSR}$ (e.g., Masson and Chernin 1992; 
Rodr{\'\i}guez et al.\ 1982; Stahler 1994; Chandler et al.\ 1996; 
Lada and Fich 1996; Gibb, personal communication). 

$$\hbox{\psfig{file=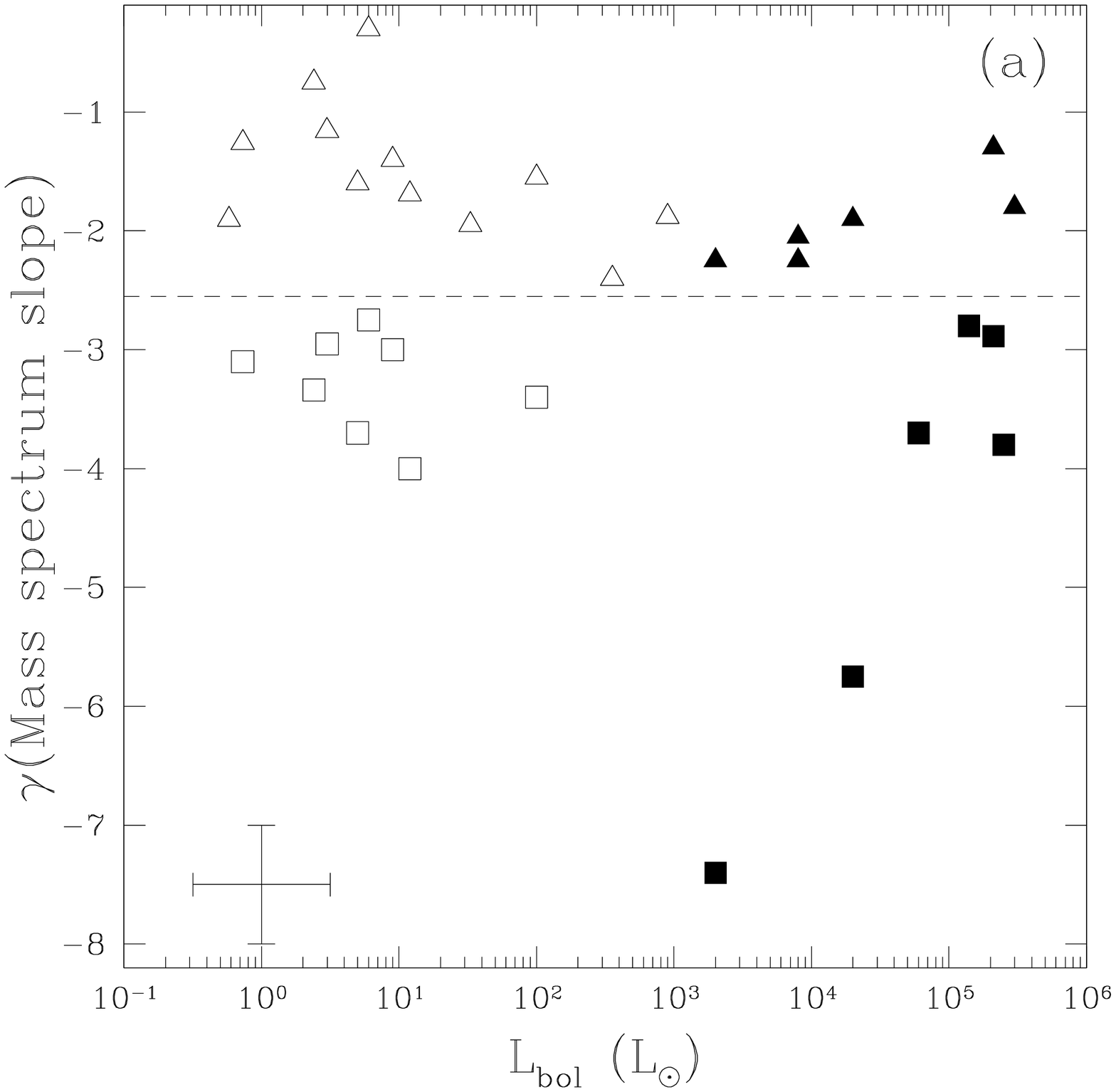,height=61mm}
\hskip1pc\psfig{file=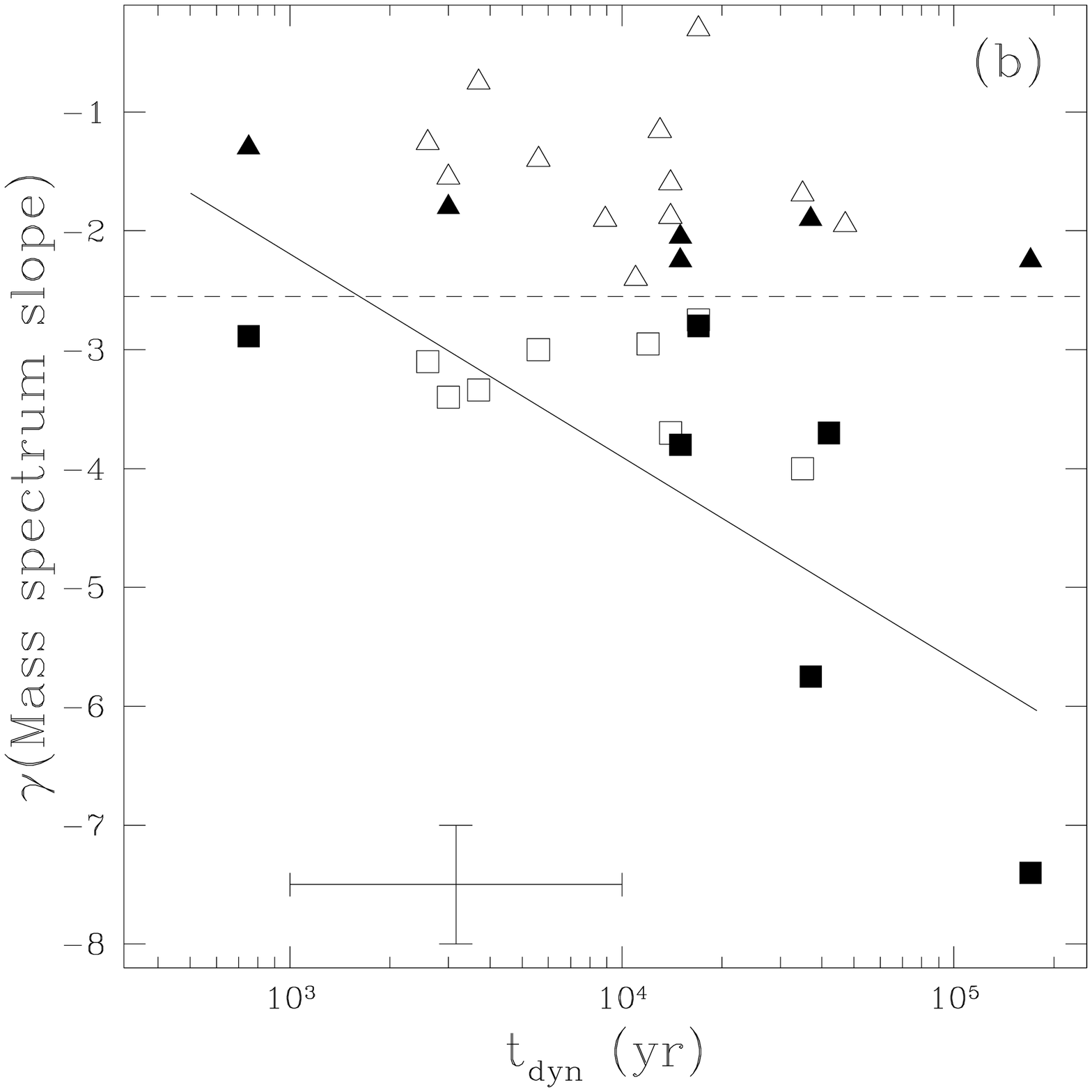,height=61mm}}$$

\caption{Figure~4.\capskip The slope $\gamma$ of the mass spectrum
$M(v)$ plotted as a function of (a) source bolometric luminosity and
(b) flow dynamical age \tdyn.  Triangles represent $\gamma$ for gas with
projected speeds less than 10~{\kms} relative to the source, and squares
are for gas moving at more than 10~{\kms}. The dashed, horizontal line
separates the low and high-velocity $\gamma$'s for clarity.  Sources with
$\Lbol < 10^3\,\Lsun$ are plotted with open symbols, those with $\Lbol >
10^3\,\Lsun$ with filled symbols. Representative error bars are displayed
in the lower left corner of each plot. The solid
line in Fig.\ 4b is a linear least squares fit (slope $-1.7 \pm 0.6$)
to high-velocity $\gamma$'s in luminous sources versus log($t_{dyn}$).
The sources plotted here are 
VLA1623, IRAS03282, L1448-C, L1551-IRS5, NGC2071-IRS1, 
and Ori~A~IRC2 (Cabrit and Bertout 1992); 
TMC-1 and TMC-1A (Chandler et al.\ 1996); 
L379-IRS1-S (Kelly and MacDonald 1996); 
NGC2264G (Lada and Fich 1996); 
G5.89-0.39 (Acord et al. 1997). 
CephE (Smith et al.\ 1997) 
HH251-254, NGC7538-IRS9, W75N-IRS1 and NGC7538-IRS1 (Davis et al.\ 1998);
G192.16 (Shepherd et al.\ 1998); 
and HH26IR, LBS17-H, G35.2-0.74, HH25MMS (Gibb, personal communication).}

Figure 4 plots (a) $\gamma$ versus $\Lbol$ and (b)
$\gamma$ versus the dynamical time scale $t_{dyn}$ for a new compilation
of 22 sources with luminosities ranging from 0.58 L$_\odot$ to $3
\times 10^5$ L$_\odot$.  Triangles represent slopes derived from gas moving
less than 10~{\kms} relative to $v_{LSR}$ while squares represent
slopes derived from gas moving more than 10~{\kms} relative to
$v_{LSR}$. Sources with $\Lbol < 10^3\,\Lsun$ are plotted with open symbols
while those with $\Lbol > 10^3\,\Lsun$ are plotted with filled
symbols. Both well-collimated and poorly-collimated outflows are
represented in the sample.

The most striking result from Fig.\ 4a is that $\gamma$'s for low-velocity
gas are similar in sources of all luminosities.  This suggests that a
common gas acceleration mechanism may operate over nearly six decades in
\Lbol.   In addition, there is a clear separation between
$\gamma$'s in high and low-velocity gas which supports the
interpretation that there are often two distinct outflow velocity
components, perhaps corresponding to a recently accelerated component
and a slower, coasting component. 

Hydrodynamic simulations of jet-driven outflows 
from low-lu\-m\-in\-os\-ity YSOs
predict such a change of slope at high velocity. They also predict that
$\gamma$ should steepen over time possibly due to the
collection of a reservoir of low-velocity gas (Smith et al.\ 1997).  Figure
4b reveals marginal evidence that the mass spectrum in 
flows from luminous YSOs does becomes
steeper with time, both in the low and high velocity ranges. The solid
line in Fig.\ 4b is a linear least squares fit (slope $-1.7 \pm 0.6$)
to high-velocity $\gamma$'s in luminous sources versus
log($t_{dyn}$). 
There is no indication of time evolution of the mass spectrum slopes in
outflows from low-luminosity sources.

%The errors and/or intrinsic variations in the
%velocity field are sufficiently large that time evolution could exist
%but we cannot identify the effect here.  

The decrease of $\gamma$ with time in more luminous sources may be due to a
difference in the driving mechanism, or it may simply be more prominent
because the mass outflow rate is several orders of magnitude greater than
in outflows from low-luminosity YSOs (thus allowing more precise
determination of $\gamma$) and because their flow ages cover a broader
range, from 750 to $2\times 10^5$ yrs.

% Hence, the effect is more readily observed.

\mainsection{{I}{I}{I}.~~TESTS OF PROTOSTELLAR WIND AND ACCRETION MODELS}
\backup
\subsection{A.~~Flow Energetics}

It is well known that outflow energetics correlate reasonably well with
\Lbol\ over the entire observed luminosity range. In Fig.\ 5a, we show a
recent compilation of the mean momentum deposition rate \FCO\ as a function
of bolometric luminosity of the driving star. It must be remembered that
\FCO\ is the time-averaged force required to drive the CO outflow:
$\FCO=\MCO\,\vCO/\tdyn$, where we assume $\tdyn$ is a good approximation to
the flow age.  If the CO-emitting material is accelerated by a separate
stellar wind (or jet), the wind momentum flux \Fw\ may be quite
different from \FCO, depending on the nature of the wind-cloud
interaction.  There are clearly large uncertainties in the measured flow
properties, primarily due to difficulties in estimating the inclination
angle of the outflow, the optical depth and excitation of the CO, and the
correctness of the assumption that the dynamical timescale is a good
estimate of the flow age (Cabrit and Bertout 1992; Padman et al.\
1997).  However, as seen in Fig.\ 5a, the correlations are roughly
consistent with a single power law (here of slope $\sim$ 0.7) across
the full range of source luminosities.  It has been suggested that this
correlation argues for a common entrainment and/or driving mechanism for
molecular flows of all masses; but this is by no means a compelling
argument, given that we would surely expect most physically reasonable
outflow mechanisms to generate more powerful winds if the source mass and
luminosity are increased.

Regardless of the details of how the stellar wind or jet entrains material,
if the shock cooling times are short compared to flow dynamical timescales
(as we expect for wind speeds less than 300\,\kms: Dyson [1984], or
if there is efficient mixing at the wind/molecular gas interface: Shu et
al.\ [1991]), then the wind and molecular flow momenta will be equal. This
is often called the momentum-conserving limit. Then, molecular
outflows represent a good opportunity to test proposed protostellar
ejection mechanisms. In particular we show in this section that they can be
used to estimate the ratio $f$ of ejection to accretion rates that would be
necessary if the wind is accretion-powered.

We first recall that optically thin,
line-driven radiative winds (such as those present in main-sequence O and B
stars) are insufficient to drive the flows if the flows are momentum
driven.  The solid line in Fig.\ 5a shows the maximum
momentum flux available in stellar photons
in the single-scattering limit, 
$\Lbol/c$. It falls short of the
observed amount in molecular flows by one to three 
orders of magnitude. If flow lifetimes have been
underestimated, the discrepancy is reduced, but not by a sufficient amount.
%Churchwell (1997a, 1997b) summarized this argument, and pointed
%out that although high optical depths in the outflow could increase the
%flow momentum by allowing multiple scattering, the scattering geometry
%required is rather unlikely.
%SC: this geometrical argument does not work: the photon can undergo
%a lot of scatterings even when staying close to the radial direction.
In principle,
higher momentum flux rates could be reached with multiple scattering.
Wolf-Rayet stars ($L_{\rm bol} \sim$ a few $10^5$ \Lsun) have a wind force
reaching $20-50L_{\rm bol}/c$, similar to the momentum rate in molecular flows
from sources of comparable luminosity (cf. Fig. 5a). Such high values are
attributed to multiple scattering of each photon by many lines closely
spaced in frequency (e.g. Gayley et al.\ 1995 and refs. therein). However,
excitation conditions in protostellar winds are very different from those
in hot, ionized Wolf-Rayet winds. The dashed line in Fig.\ 5a shows the
typical momentum flux in the {\it ionized} component of protostellar winds,
inferred from recombination lines or radio continuum data (Panagia 1991):
the values lie a factor of 10 below \FCO, implying that the driving winds
must be 90\% 
neutral. 
If dust grains instead provide the
dominant opacity source in protostellar winds, comparison with winds
from cool giants and supergiants might be more appropriate. Their
wind velocities are $\simeq$ 5-30 km s$^{-1}$
and mass-loss rates do not exceed 10$^{-4}$ \Msun\ \yr, hence $F_w \le
0.5-3L_{\rm bol}/c$ for a typical $L_{\rm bol} \sim 5\times 10^4$
\Lsun. Calculations by Netzer and Elitzur (1993) show that 10 times larger
mass-loss rates could in principle be achieved in oxygen-rich stars,
where silicates dominate the opacity curve. The wind force could then
become comparable to the molecular outflow momentum rate for $L_{\rm
bol} \sim 5\times 10^4\,\Lsun$.  Hence it is just possible that
luminous stars ($\Lstar>5\times10^4\,\Lsun$) could drive their flows
by radiative acceleration if dust opacity plays a significant role;
further work is needed to investigate if such models are viable. In
lower luminosity sources, however, the opacity required to lift
material above escape speeds largely exceeds typical values for
circumstellar dust.  Therefore, molecular outflow sources with
$\Lstar<5\times10^4\Lsun$ must possess an efficient non-radiative
outward momentum source.

%Eventually, however, opacity will stop increasing when radiation is
%degraded to wavelengths longer than the silicate bands. Therefore it seems
%highly unlikely that radiation driving on dust could explain forces of
%100-1000$L_{\rm bol}/c$ observed in flows from lower luminosity sources. In
%addition, gravity is much higher in accreting protostars than in giants and
%supergiants, thus presumably lowering the radiation driving efficiency.

It thus appears most likely that both low and high-mass systems
possess an efficient non-radiative wind-generation mechanism in their
embedded protostellar phase.  If energetic bipolar winds are the chief
means of angular momentum loss during the main accretion phase for
stars of all mass, this is not surprising. However, the details of the
wind ejection mechanism could differ; in particular massive accreting
stars are likely to have thinner convective layers and probably rotate
faster, so that the magnetic field and accretion geometry close to the
star may be very different from that in low-mass stars.

There exist quite a number of efficient accretion-powered wind mechanisms
from the stellar surface, disk, or disk-mag\-ne\-to\-sphere bou\-nd\-ary.
The most efficient models use a
strong magnetic field in the star or disk to drive the wind and to carry
off angular momentum from the accreting gas (see e.g. reviews by
K\"onigl and Pudritz, and Shu et al., this volume). This wind is further
collimated by magnetic or hydrodynamic processes (e.g. Mellema et al.\
1997), generating a high-Mach number wind and/or jet with a speed of order
200-800\,\kms. A fraction $f$ $<1$ of the accretion flow \Madot\ is ejected
in the wind: $\Mwdot = f\,\Madot$.

A different scenario recently explored by Fiege and Henriksen (1996a,
1996b) is that molecular flows are not predominantly swept-up by an
underlying wind, but represent infalling gas that has been deflected
into polar streams by magnetic forces. Only a small fraction of the
infalling gas actually reaches the star to produce accretion
luminosity. In that case, $f > 1$. This model has been invoked in
particular to explain the large flow masses $\MCO \sim 10M_*$ observed
in flows from luminous sources.  In the following we use molecular
flow observations to set constraints on these two classes of proposed
models.

\subsection{B.~~The Ejection/Accretion Ratio in Protostellar Winds}

Two simple, independent methods have been used in the literature to
estimate $f$, assuming that the driving mechanism is steady over the
source lifetime. First, in low-mass protostars where the luminosity is
accretion-dominated, it is possible to use the observed correlation of
flow force with \Lbol\ (Fig.\ 5a) to derive the ejection fraction
$f$. The source bolometric luminosity is
$\Lbol=G\Mstar\Madot/\Rstar=\Madot\,\vk^2$, where
\vk\ is the Keplerian speed at the stellar surface. The flow and wind
force (assuming momentum conservation) is $\FCO = f\,\Madot\,\vw$. Hence
$\FCO/\Lbol = f\,\vw/\vk^2$. A value $f
\sim 0.1$ is inferred in both Class 0 and Class I low-luminosity
objects (Bontemps et al.\ 1996). Alternatively, if one estimates
\Mstar\ from \Lbol\ via the ZAMS relationship (which is probably only
valid for sources with \Lbol $> 10^3$ \Lsun), one can use the {\it
accumulated} flow momentum to infer $f$, using: $\PCO = \vw \, \Mw = f
\, \Mstar\, \vw$.  Values of $f$ ranging from 0.1-1 are inferred for a
wind speed of 150 \kms\ (Masson and Chernin 1994; Shepherd et al.\
1996). However, we point out that wind speeds are unlikely to remain
constant over the whole \Lbol\ range.  There is evidence for higher wind
velocities in luminous flow sources: for example, proper motions up to
1400\,\kms\ are seen in HH80-81 (Mart\'{\i} et al.\ 1995) and Z~CMa shows
optical jet emission with speeds up to 650\,\kms\ (Poetzel et al. 1989).  These
are significantly higher than the 100-200\,\kms\ typically seen in low-mass
sources.

To re-examine this issue in a homogeneous way over the whole luminosity
range, we plot in Fig.\ 5b the values of $f\vw/\vk$
obtained using a new combination of the above two methods (Cabrit and
Shepherd, in preparation). The solid circles are derived assuming the
luminosity is accretion dominated, while the open circles (for sources more
luminous than $10^3\,\Lsun$) assume the ZAMS relationship given above.
Typically, \vk\ ranges from 100\,\kms\ in low-luminosity sources to
800\,\kms\ in high-luminosity sources. A rather constant value $f
\vw/\vk \sim 0.3$ seems to hold over the whole range of
\Lbol.  This value is in line with both popular MHD ejection models.
In the X-wind model (Shu et al.\ 1994, and this volume), the wind is
launched close to the stellar surface ($\vw \sim \vk$) and a large
fraction of the accreting gas is ejected ($f \sim 0.3$).  In the
self-similar disk-wind models (Ferreira 1997; 
K\"onigl and Pudritz, this volume),
less material is ejected ($f \sim 0.03$) but the long magnetic lever
arm accelerates it to many times the Keplerian speed ($\vw \sim
10\vk$). Thus we conclude that the {\it energetics} of the flows over
the entire luminosity range are broadly consistent with a unified MHD
ejection model for all flow luminosities, but we reiterate that given
the very different physics involved around high and low-mass
protostars, the details of such a model for high-mass systems are
still unclear.

$$\psfig{file=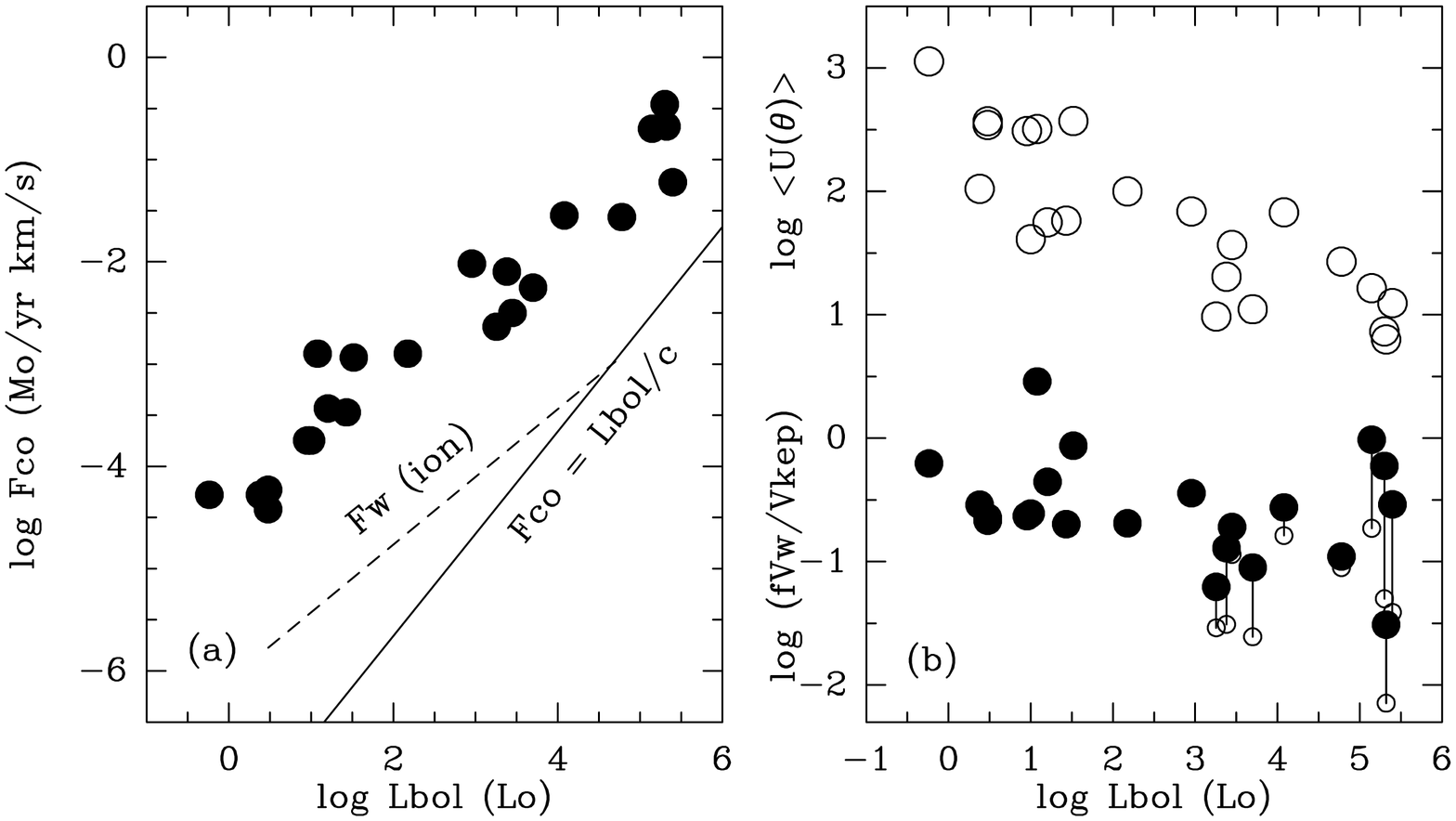,angle=270,height=99mm}$$
\caption{Figure~5.\capskip (a) The average momentum flux in the CO
flows as a function of source luminosity. The solid line shows the
force available in stellar photons (assuming single scattering), and
the dashed line the force available from the ionized wind components
(Panagia 1991).  (b) The factor $f\vw/\vk$ derived for the same set of
sources is plotted as solid circles (small open circles assume that
luminous sources are on the ZAMS). Large open circles show the
required $\langle U(\theta)\rangle$ factors for the same objects if
the circulation models are applicable (see text for details). }

We stress again that this plot assumes perfect momentum conservation
in the wind/flow interaction. There could be strong deviations from
this key assumption. First, we should keep in mind the possibility
that massive flows enter the energy-driven regime: for high wind
speeds, the gas cooling behind the shock will be slow, and the
snowplow or momentum-conserving flow will turn into an energy driven
one. In this case, the shell momentum can exceed the momentum in the
wind itself by a factor of order $\vw/\vCO$ (Cabrit and Bertout 1992;
Dyson 1994) so reducing markedly the momentum requirement of
the driving wind. Of course, there are objections to energy driven
flows, in particular their inability to reproduce the bipolar velocity
fields of many flows (Masson and Chernin 1992); but given the
apparently poorer collimation of high-mass outflows this issue should
perhaps be re-examined.  Second, if the flow is entrained in a jet
bow-shock, the efficiency of momentum transfer will depend on the
ratio of ambient to jet density: it will only be close to 1 if the jet
is less dense than the ambient medium.  Highly overdense jets will
pierce through the cloud without depositing much of their momentum
(Chernin et al.\ 1994), and in that case the ratio $f\vw/\vk$ plotted
in Fig.\ 5b would have to be increased.  We conclude that for current
protostellar jet models to apply, we have the additional condition
that most of the jet momentum must be in a component that is not
significantly denser than the ambient molecular cloud on scales of
0.1-1pc.

The above analysis also provides important constraints on accretion rates
in protostellar objects of various masses: if $f\vw/\vk \sim 0.3$, then the
values of \FCO\ in Fig.\ 5a show that $\Madot = 3\FCO/\vk$ must range from a
few $10^{-6}$ \Msun \yr at \Lbol $\sim$ 1\Lsun\ to a few $10^{-3}$ \Msun
\yr at \Lbol $\sim 10^5$ \Lsun. Then, protostellar sources would not be
characterized by a single infall rate across the whole stellar mass range,
and massive stars would form with much higher infall rates than low-mass
stars (Cabrit and Shepherd, in preparation).

The wealth of data available on low-mass systems also allows one to break
the samples down by estimated age, and so look for {\it evolution} of
outflow properties with stellar age. Bontemps et al.\ (1996) made an
important study of low-mass systems in Taurus and Ophiuchus, and found
evidence for a secular decline in outflow power with age, while $f$
remained constant. Intriguingly, the best correlation of outflow power
was with circumstellar mass (as measured by the millimeter continuum
flux) rather than with source bolometric luminosity; Saraceno et al
(1996) also presented a similar correlation between millimeter
continuum flux and outflow kinetic luminosity in systems with
$L<10^3\,\Lsun$. These results strongly suggest that the accretion
rate and the outflow strength both decline in proportion to the disk
and envelope mass.

\subsection{C.~~Deflected Infall Models}

Although the above analysis shows
that $f$ need not be necessarily higher in high-mass outflows, the
very large masses involved, combined with the inefficiency of entrainment
and momentum transfer by dense jets (especially once they escape their
parent clouds), has led to some discussion of whether the sweeping up
of ambient molecular gas by an accretion driven wind is a viable
mechanism for these objects (e.g. Churchwell 1997a).

A recent class of outflow models, termed circulation models, can naturally
generate outflow masses much greater than the stellar mass.  In these, most
of the infalling circumstellar material is diverted magnetically at large
radii into a slow-moving outflow along the polar direction, while infall
proceeds along the equatorial plane (Fiege and Henriksen 1996a, 1996b). The
main attraction of these models is that they generate large outflow masses
for even small stellar masses and can generally explain the observed
opening angles and velocity structure seen in high-mass systems.  In
particular, self-similar models predict that the velocity and density
laws should take the form $V(r,\theta) = U(\theta)
\sqrt{GM_*/r}$ and $\rho(r,\theta) =
\mu(\theta) M_* r_o^{-3} (r/r_o)^{2\alpha-0.5}$, where
$0.25 \ge \alpha > -0.5$, $r_o$ is an unspecified radial scale, and
$U(\theta)$ and $\mu(\theta)$ are dimensionless functions. It is then
straightforward to show that the force and mass-flux in the outflow are
related by $\FCO/\dot{M}_{CO} = \sqrt{G\Mstar/R_{CO}}\langle
U(\theta)\rangle$ where $\langle U(\theta) \rangle$ = $\int{U(\theta)^2
\mu(\theta) d\omega} / \int{U(\theta) \mu(\theta) d\omega}$ is the
density-weighted average velocity over the outflow solid angle.  The
inferred $\langle U(\theta)\rangle$ (using the same \Mstar\ as for our
estimates of $f \, \vw / \vk$) is plotted in the top part of Fig.\ 5b as open
circles. It is clear that high values (between 1000 and 10) are required. 
Generalized circulation models that include Poynting
flux driving yield values of $\langle U(\theta) \rangle$ between 5 and 200
(Lery, Henriksen, Fiege, in preparation).  
In model cases where radiation transport is important in setting up
the flow, a power-law slope of 0.8 is predicted between $F_{CO}$ and \Lbol,
which is close to the observed slope $\sim 0.7$ (see also Henriksen
1994). Observations seem to indicate a systematic decline of $\langle
U(\theta) \rangle$ with \Lbol, which would also have to be explained.

There are several concerns about these circulation models. First,
there are many {\it unipolar} CO outflows known, such as NGC2024-FIR5
(Richer et al.\ 1992), and the almost-unipolar HH46-47 system (Chernin and
Masson 1991). These are naturally explained by swept-up wind models if the
protostar is forming on the edge of a cloud or close to an HII region
interface: the jet or wind propagating into the cloud will then sweep up a
large CO flow, while in the opposite direction little evidence for a CO
lobe will be seen. 
In circulation
models, anisotropic solutions may also occur, but it is unclear why the
weaker lobe would necessarily be on the side where the large-scale cloud
density is low
Second, in some objects such as B5 (Velusamy and
Langer 1998), there is an apparent lack of molecular material in the
equatorial plane which can feed a circulation flow. Third, as
discussed in section II, in some high-mass systems such as HH80-81 there is
direct evidence for fast jets and bow-shock entrainment of molecular
gas. Consequently, it seems more probable given the evidence presented
that even high-mass outflows can be generated by the sweeping up of ambient
cloud material by an accretion-driven stellar wind or jet. The details of
the MHD driving mechanism in these cases, and of the momentum transfer
between the wind and the jet, remain open issues.

\mainsection{{I}{V}.~~SHOCK CHEMISTRY AND ENERGETICS}

The interaction between a supersonic protostellar wind and surrounding
quiescent material is expected to drive strong shock fronts.  Shocks can be
of type C (continuous) or J (jump) depending upon the shock velocity, the
magnetic field, and the ionization fraction of the pre-shock gas (Draine
and McKee 1993; Hollenbach 1997). In the last few years, spectacular gains
in sensitivity in the mm and IR domains have allowed us for the first time
to witness the chemical and thermal effects of these shocks in molecular
outflows. These observations yield direct estimates of the flow
age, energetics, and entrainment conditions which represent an important 
new step toward a complete description of the outflow phenomenon. 

\subsection{A.~~Chemical Processing of ISM in Molecular Flows}

{\it Theoretical expectations:} By compressing and heating the gas, shock
waves trigger new chemical processes which lead to a specific ``shock
chemistry'' (see chapter by Langer et al., this volume). 
The most active molecular chemistry is expected to occur in
C-shocks, as they increase the temperature to moderate values of about 2000
K in a thick layer where molecules can survive, and reactions that
overcome energy barriers can proceed. In particular, the very reactive
OH radical can be formed by O+H$_2$ $\to$ OH + H (which has an energy
barrier of 3160 K), and will contribute to the formation of H$_2$O by
further reaction with H$_2$: OH + H$_2$ $\to$ H$_2$O + H (energy barrier:
1660 K). In dissociative J-shocks, molecules are destroyed in the hot (T
$\sim$ 10$^5$ K) thin post-shock layer and only reform over longer
timescales, in a plateau of gas at $\sim$ 400 K.  Since some of these
chemical processes are fast, and the cooling times are short, the chemical
composition of the shocked regions is expected to be strongly
time-dependent.

Shocks also process dust grains. In the most violent J shocks,
destruction of grain cores and thermal sputtering inject refractory
elements (such as Si and Fe) into the gas phase (e.g. Flower et al.\ 1996). 
In slower C-type shocks,
non-thermal sputtering will inject refractory and volatile species mainly
from the grain mantles into the gas phase (e.g. Flower and Pineau 
des For\^ets 1994).  The entrance of this fresh
material, together with the high abundance of OH, will
produce oxides such as SO and SiO (see Bachiller 1996, van
Dishoeck and Blake 1998, and references there in). As the shocked gas
cools, the dominant reactions will be again those of the usual low
temperature chemistry, and depletion onto dust grain surfaces will reduce
the abundances of some of the newly formed molecules (e.g. H$_2$O, see
Bergin et al.\ 1998).  However, the chemical composition of both the gas
and the solid phases will remain altered with respect to pre-shock ones.

{\it Observations:} The chemical effects of shocks have been observed
in a number of outflows from low-mass Class 0 objects, which are
particularly energetic and contain shocked regions well separated
spatially from the quiescent protostellar envelope.  Recent examples
include IRAS16293 (Blake et al.\ 1994; van Dishoeck et al.\ 1995),
NGC1333 IRAS4 (Blake et al.\ 1995), and NGC1333 IRAS2 
(Langer et al.\ 1996; Blake 1997; Bachiller et al.\ 1998). 

A comprehensive study of many
different species has been recently carried out on L1157 (Bachiller and
P\'erez Guti\'errez 1997).  The abundances of many molecules
(e.g. CH$_3$OH, H$_2$CO, HCO$^+$, NH$_3$, HCN, HNC, CN, CS, SO,
SO$_2$) are observed to be enhanced by factors ranging from a few to a
few hundred. The extreme case is SiO which is enhanced by a factor of
$\sim 10^6$.  
There are significant differences in spatial distribution among the
different species: some molecules such as HCO$^+$ and CN peak close to the
central source, while SO and SO$_2$ have a maximum in the more distant
shocks, with OCS having the most distant peak.  Other molecules such as
SiO, CS, CH$_3$OH, and H$_2$CO show an intermediate behavior. Such
differences cannot be attributed solely to excitation conditions: an
important gradient in chemical composition is observed along the
outflow. It is very likely that this strong gradient is related to the time
dependence of shock-chemistry. As an example, consider the chemistry of SO,
SO$_2$, and OCS, which has been recently modeled by Charnley (1997). It is
believed that sulfur is released from grains in the form of H$_2$S, and
that it is then oxidized to SO and SO$_2$ in a few 10$^3$ yr. The formation
of OCS needs a few 10$^4$ yr.  This is in general agreement with
observations, since the SO/H$_2$S and SO$_2$/H$_2$S ratios do increase with
distance from the source (i.e. with time), and OCS emission is only
observed in the most distant position (i.e the oldest shock). Hence,
chemical studies are of high potential to constrain the age and time
evolution of molecular outflows.

\subsection{B.~~Shock Cooling and Energetics}

Millimeter observations of sh\-ock-en\-hanced mo\-le\-cules trace
chemically processed gas  that has already cooled
down to 60-100 K, as indicated e.g. by multi-line NH$_3$ studies (Bachiller
et al.\ 1993, Tafalla and Bachiller 1995). Emission from hotter post-shock
gas, on the other hand, is important to obtain information on
the instantaneous energy input rate and pre-shock conditions in outflows.

{\it Hot (T $\ge$ 1000 K) shocked gas:} Hollenbach (1985) suggested that
the [O~I]\,63\mi\ line should offer a useful, extinction-insensitive measure
of {\it dissociative} J-shocks in molecular flows. Because [O~I]\,63\mi\ is
the main coolant below $\sim$ 5000 K, its intensity is roughly proportional
to the mass flux {\it into the J-shock}, $\dot{M}_{JS}$, through the
relation $L_{[O I]}/L_\odot = 10^{4} \times
\dot{M}_{JS} / M_\odot {\rm yr}^{-1}$, as long as the line remains optically
thin (i.e. $n_o V_{JS} < 10^7$ \kms\ \cm3, where $n_o$ is the pre-shock
density and $V_{JS}$ is the J-shock velocity).

First detections of [O~I]\,63\mi\ in outflows were obtained with the Kuiper
Airborne Observatory (KAO) toward 3 HH objects and 5 highly collimated Class
0 outflows (Cohen et al. 1988; Ceccarelli et al. 1997). Since the advent of
the Infrared Space Observatory (ISO), the Long Wavelength Spectrometer
(LWS) has revealed [O~I]\,63\mi\ emission in at least 10 more HH objects and
molecular outflows (Liseau et al. 1997; Saraceno et al. 1998).  These
authors find a surprisingly good correlation between {\it current} values of
$\dot{M}_{JS}$ derived from [O~I]\,63\mi\ and {\it time-averaged}
$\dot{M}_{ave}$ values derived from mm observations of the outflow assuming
ram pressure equilibrium at the shock (i.e. \FCO = $\dot{M}_{ave}\times
V_{JS}$).  Both values agree for $V_{JS} \sim$ 100 \kms. The dispersion in
this correlation, roughly a factor of 3, is of the same order as the
uncertainties in \FCO\ caused by opacity and projection effects (Cabrit and
Bertout 1992). Hence, CO-derived momentum rates in outflows do not appear
to suffer from large systematic errors.

With the development of large format near-IR arrays in the early 1990's, it
has also become possible to map molecular outflows in the 2.12\mi\ $v =
1-0$ S(1) line of \Htwo, a tracer of hot ($\sim$ 2000 K) shocked molecular
gas.  In both low-luminosity and high-luminosity outflows, the \Htwo\
emission delineates single or multiple bow-shaped features associated with
the leading edge of the CO emission (Davis and Eisl\"offel 1995; Davis et
al.\ 1998), as illustrated in Fig.\ 1 for HH211; in a few cases, \Htwo\
emission also traces collimated jets and cavity walls (e.g Bally et al.\
1993; Eisl\"offel et al.\ 1994).  Observed surface
brightnesses and rotational temperatures $\sim$ 1500-2500 K indicate
moderate velocity shocks: either J-shocks of speed 10-25 \kms\ or C-shocks
with $V_s \sim 30$ \kms\ and low filling-factor (Smith 1994; Gredel
1994). In particular, the morphology, line profile shapes, intensity, and
proper motions of \Htwo\ 2.12$\mu$m bows are well reproduced by
hydrodynamical simulations of jets propagating into the surrounding cloud,
where \Htwo\ 2.12$\mu$m emission arises mostly in the non-dissociative
wings of the bowshock 
(Raga et al.\ 1995; Micono et al.\ 1998; Suttner et al.\ 1998).

If these bowshocks are also where most of the slow molecular outflow is
being accelerated, and if \Htwo\ emission dominates the cooling (as
expected e.g. in 2000K molecular gas at densities of $10^5-10^8$ \cm3),
then $L$(\Htwo)$/\LCO$ should be of order unity (see e.g. Hollenbach
1997). In the 5 flows studied by Davis and Eisl\"offel (1995), the observed
ratio $L(\Htwo)/\LCO$ has a median value of $\sim$0.4, but it covers a very
broad range from 0.001 to 30. The discrepancies could be caused by
uncertainties in 2\mi\ extinction (corrections typically amount to 10-100
for $A_V = 20-50$ mag), by the use of unreliable
\LCO\ estimates, or --- in the case of very low ratios --- by a strong
decrease in outflow power over time (see W75N; Davis et al. 1998). Hence
the \Htwo\ 2.12\mi\ line alone is not a sufficient
diagnostic of the outflow entrainment process.

{\it Warm (T $\sim$ 300-1000 K) molecular gas}:
The ISO mission has led to the detection of a new component of warm
post-shock gas at T $\sim$ 300-1000 K in several outflows. Fig.\ 6
shows a map of the L1157 outflow in the $v = 0-0$ S(5) pure rotational
line of \Htwo\ at 6.9\mi\ obtained with ISOCAM (Cabrit et al.\ 1998). A
series of bright emission spots are seen along the outflow axis. They
coincide spatially with hot shocked gas emitting in the
\Htwo\ 2.12\mi\ line (Eisl\"offel and Davis 1995), and with the various
peaks of shock-en\-han\-ced molecules identified by Bachiller and P\'erez
Guti\'errez (1997; see Sect. IV.A). However, they trace an intermediate
temperature regime of $\sim$ 800 K, considerably lower than the 2000 K
observed in ro-vibrational \Htwo\ lines. Warm \Htwo\ at 700-800K was also
found with the ISO Short Wavelength Spectrometer (SWS) in two outflows from
very luminous sources, Cep~A and DR~21 (Wright et al.\ 1996; Smith et al.\
1998). Finally, warm CO at $T \sim 330-1600 K$ was detected in high-J lines
($J_{up}$ = 14 to 28) with ISO-LWS toward 5 outflows of various luminosities,
while H$_2$O and OH lines were detected in two cases (Nisini et al.\ 1996,
1997; Ceccarelli et al.\ 1998).

%It is clear that emission cannot arise in reformed molecules behind a {\it
%dissociative} J-shock (see e.g. Ceccarelli et al. 1998): (1) LVG studies of
%the CO lines indicate moderate densities in the emitting region $\sim
%10^4-10^5$ \cm3, hence low pre-shock densities $\le 10^4$ which do not lead
%to substantial molecular column densities behind a dissociative J-shock;
%(2) The observed typical temperature of 700K is also incompatible with a
%molecular reformation plateau behind a J-shock ($T \sim 400 K$; Hollenbach
%and McKee 1989); (3) Finally, CO emission is much stronger than [O~I]63\mi.

The observed emission fluxes and temperatures in \Htwo\ and CO are well
explained by non-dissociative J-shocks with $V_s \sim$ 10 \kms\ or slow
C-shocks with $V_s \sim$ 10-25 \kms, and $n_o \sim 10^4-3\times 10^5$ \cm3\
(e.g. Wright et al. 1996; Nisini et al. 1997; Cabrit et al. 1998).  One
important constraint is the rather low [H$_2$O]/[\Htwo] abundance ratio $\sim
1-2\times 10^{-5}$ observed in HH54 and IRAS16293 (Liseau et al.\ 1996;
Ceccarelli et al. 1998); steady-state C-shocks would predict complete
conversion of O into water. The relatively high [OH]/[H$_2$O] ratio $\sim$
1/4-1/10 in these two flows suggests that the shock age is too short for
conversion to be complete, and points to the need for time-dependent
C-shock models for proper interpretation of the data (e.g. Chi\`eze et al.\
1998). 

Mid and far-infrared emission from this warm molecular gas component
appears more tightly correlated with \LCO\ than the 2\mi\ \Htwo\ lines.  In
4 out of the 5 outflows studied by Nisini et al.\ (1997), the FIR CO
luminosity represents 10-30\% of the flow kinetic luminosity, in good
agreement with C-shock calculations in the inferred density and velocity
range (Kaufman and Neufeld 1996). The only large discrepancy is observed in
IC1396N, an object contaminated by PDR emission (Molinari et al.\ 1997). In
L1157, the \Htwo\ luminosity of warm gas is also around 10\% of
\LCO\ (Cabrit et al., 1999). Hence, these slow shocks seem sufficient to
drive the whole outflows. Detailed comparisons between \Htwo\ and FIR-CO
lines in the same objects are now under way to further narrow the
range of possible shock models and perhaps allow us to discriminate between
wide-angle wind and jet scenarios for the entrainment of outflows.

$$\psfig{file=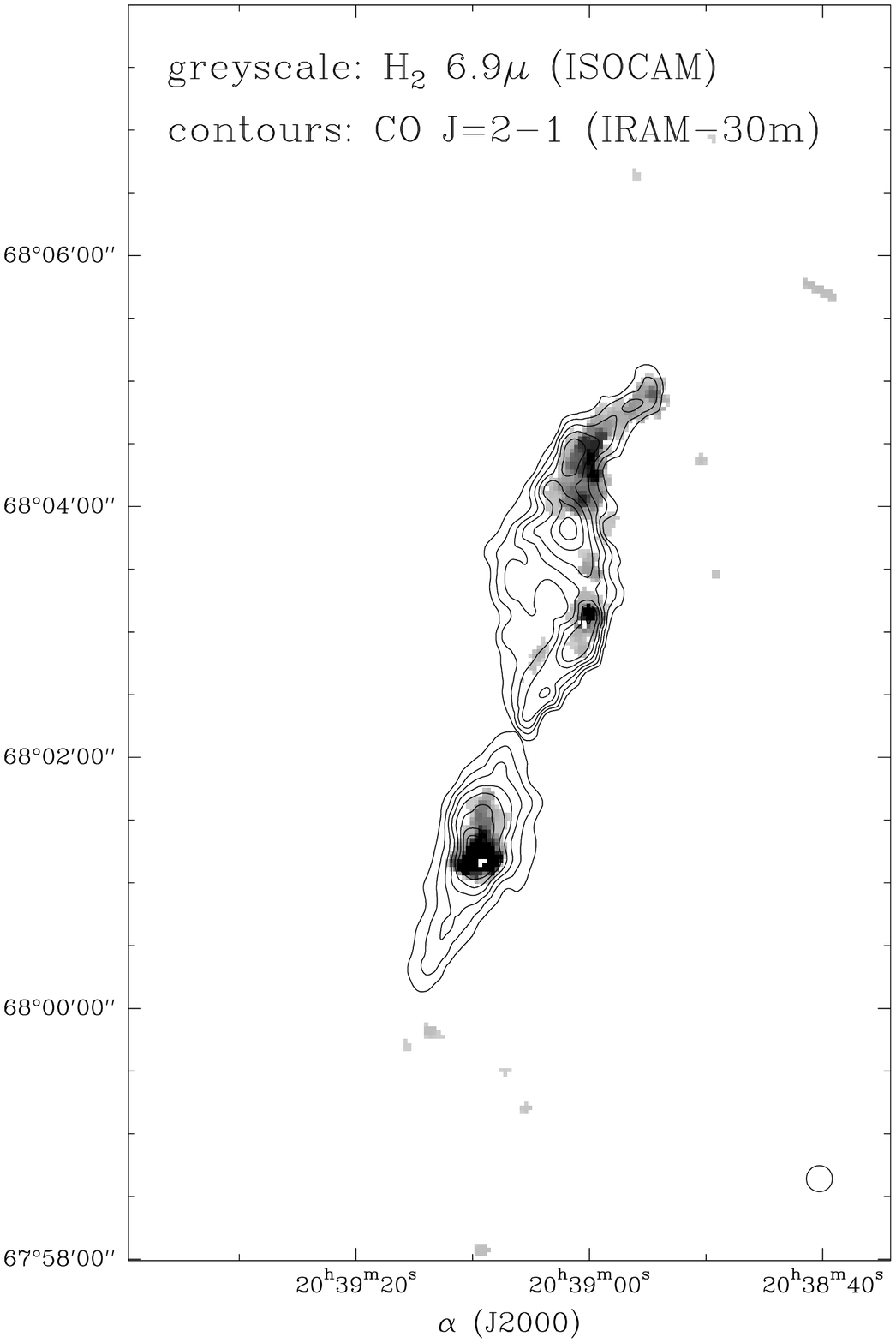,height=140mm}$$
\caption{Figure~6.\capskip
%(Left) maps of the southern lobe of L1157 in various
%molecular species, showing differential chemical processing along the
%outflow (from Bachiller and P\'erez Gutierrez, 1997). (Right) 
The L1157 outflow mapped in the 6.9\mi\ pure rotational line of \Htwo\
(adapted from Cabrit et al.\ 1998) with CO(2-1) contours
superimposed (from Bachiller and P\'erez Guti\'errez 1997).}

\mainsection{{V}~~CONCLUSIONS}

We have shown that the current data on the structure and energetics of
molecular outflows suggest broad similarities across the entire
luminosity range, from 1 to $10^5\,\Lsun$. If the flows are swept up
by a stellar wind or jet, we find that $\Mwdot\vw/\Madot\vk$ has a
value of about 0.3 for all flows, perhaps suggesting that flows have a
common drive mechanism. However, it remains unclear if the MHD disk
and X-wind models which have been used to explain low-mass outflows
are appropriate in the very different physical regime of high-mass
YSOs. While observational data continue to improve these constraints,
there remains an urgent need for a larger sample of molecular
outflows, particularly from high-mass stars, to be fully mapped at
high resolution; at the moment it is very possible that our estimates
of the properties of high-mass systems are biased by strong selection
effects. Single dish data, especially from the new focal plane arrays,
plus interferometric images at millimeter wavelengths will continue to
accumulate. However, only when the large millimeter interferometer
(MMA/LSA) is operational will it be possible to acquire
high-resolution data quickly enough to study large samples of outflows
in detail.

The nature of the shocks which drive outflows is slowly becoming
clearer. We now have diagnostics of all the temperature components in
the outflows, from the 2000K gas seen in the 2\,\micron\ \Htwo\ lines,
to the several hundred Kelvin component recently detected by ISO,
through to the cool massive component seen in the millimeter waveband
where most of the momentum is eventually deposited. The relationship
between these components is providing valuable tests of the outflow
mechanism, although a fuller understanding will require observations
at higher angular resolution than ISO provided. The SOFIA and
especially the FIRST missions will provide valuable data in this area.

\vskip 0.25in

{\bf Acknowledgments:~~} DSS would like to thank Andy Gibb and Chris
Davis for providing unpublished outflow data for a number of sources
and A. Gibb, A. Sargent, and L. Testi for useful discussions.  JSR
acknowledges support from the Royal Society.

\vfill\eject\null
\vskip .5in
\centerline{\bf REFERENCES}
\vskip .25in

\ref{Acord, J. M., Walmsley, C. M., and Churchwell, E. 1997.
The Extraordinary Outflow toward G5.89-0.39.
In {\refit Astrophy. J.\/}, 475:693--704.}

\ref{{Bachiller}, R., {Codella}, C., {Colomer}, F., {Liechti}, S.
	and {Walmsley}, C. M. 
1998. 
Methanol in protostellar outflows. Single-dish and interfero\-metric
maps of NGC 1333-IRAS~2.
{\refit Astrophys.\ J.\/} 335:266--276.}

\ref{{Bachiller}, R., {Tafalla}, M. and {Cernicharo}, J.
1994.
Successive ejection events in the L1551 molecular outflow.
{\refit Astrophys.\ J.\ Lett.\/} 425:93--.
}

\ref{Bachiller, R., Cernicharo, J., 
Mart\'{\i}n-Pintado, J., Tafalla, M., and Lazareff, B. 
1990. 
High-velocity molecular bullets in a fast bipolar outflow near L1448/IRS3. 
{\refit Astron.\ Astrophys.\/} 231:174--86.}

\ref{Bachiller, R., Martin-Pintado, J. and Fuente, A. 
1993. 
High-Velocity Hot Ammonia in Bipolar Outflows. 
{\refit Astrophys.\ J.\ Lett.\/} 417:45--48.}

\ref{{Bachiller}, R. and {P\'erez Guti\'errez}, M.
1997.
Shock Chemistry in the Young Bipolar Outflow L1157.
{\refit Astrophys.\ J.\ Lett.\/} 487:93-97.}

\ref{Bachiller, R., Guilloteau, S., Dutrey, A., Planesas, P., and Mart\'{\i}n-Pintado, J. 
1995. 
The jet-driven molecular outflow in L 1448. CO and continuum synthesis images.
{\refit Astron.\ Astrophys.\/} 299:857--.}

\ref{Bachiller, R. 
1996.
Bipolar Molecular Outflows from Young Stars and Protostars. 
{\refit Ann.\ Rev.\ Astron.\ Astrophys.\/} 34:111--154.}

\ref{Bally, J., {Devine}, D., {Hereld}, M., and  {Rauscher}, B. J.
1993.
Molecular Hydrogen in the IRAS 03282+3035 Stellar Jet.
{\refit Astrophys.\ J.\ Lett.\/} 418:75--.
}

\ref{{Bence}, S. J., {Padman}, R., {Isaak}, K. G. {Wiedner}, M. C., and {Wright}, G. S.
1998.
L~43: the late stages of a molecular outflow.
{\refit Mon.\ Not.\ Roy.\ Astron.\ Soc.\/} 299:965--
}

\ref{{Bence}, S. J., {Richer}, J. S. and {Padman}, R.
1996.
RNO~43: a jet-driven super-outflow.
{\refit Mon.\ Not.\ Roy.\ Astron.\ Soc.\/} 279: 866--883.
}

\ref{{Bergin}, E. A., {Neufeld}, D. A. and {Melnick}, G. J. 
1998. 
The Postshock Chemical Lifetimes of Outflow Tracers and a Possible New
Mechanism to Produce Water Ice Mantles.
{\refit Astrophys.\ J.\/} 499:777--.
}

\ref{{Blake}, G. A., {van Dishoeck}, E. F., {Jansen}, D. J., 
	{Groesbeck}, T. D. and {Mundy}, L. G. 
1994. 
Molecular abundances and low-mass star formation. 1: Si- and S-bearing
species toward IRAS 16293-2422. 
{\refit Astrophys.\ J.\/} 428:680--692.}

\ref{{Blake}, G. A., {Sandell}, G., {Van Dishoeck}, E. F., 
	{Groesbeck}, T. D., {Mundy}, L. G. and {Aspin}, C. 
1995. 
A molecular line study of NGC~1333/IRAS~4. 
{\refit Astrophys.\ J.\/} 441:689--701.}

\ref{Blake G.A. 
1997.
High angular resolution observations of the gas phase composition of
young stellar objects. 
In {\refit IAU Symposium No. 178\/}, ed. E. van Dishoeck, pp. 31--.}

\ref{Bontemps, S., Andr\'e, P., Terebey, S. and Cabrit, S.
1996. 
Evolution of outflow activity around low-mass young stellar objects. 
{\refit Astron.\ Astrophys.\/} 311:858--872.}

\ref{Cabrit, S. and Bertout, C. 
1992.  
CO line formation in bipolar flows. 3.~The energetics of molecular flows and ionized winds. 
{\refit Astron.\ Astrophys.\/} 261:274--284.}
% Lbol vs Lmech and F in outflows

\ref{Cabrit S., Raga, A. and Gueth, F. 
1997.  
Models of bipolar molecular outflows.  
In {\refit IAU Symposium No. 182\/}, eds.\ B. Reipurth and 
C. Bertout (Kluwer Academic Publishers), pp.\ 163--180.}

\ref{{Cabrit}, S., {Goldsmith}, P. F. and {Snell}, R. L.
1988.
Identification of RNO 43 and B335 as two highly collimated bipolar
flows oriented nearly in the plane of the sky.
{\refit Astrophys.\ J.\/} 334:196--208.
}

\ref{{Cabrit}, S.,  {Couturier}, Andre, P.,  {Boulade}, O.,  {Cesarsky}, C. J.
	and {Lagage}, P. O.,  {Sauvage}, M.,  {Bontemps}, S.,  {Nordh}, L.
	and {Olofsson}, G.,  {Boulanger}, F.,  {Sibille}, F., and {Siebenmorgen}, R.
1998. 
Mid-Infrared Emission Maps of Bipolar Outflows with ISOCAM: an in-depth study of the L1157 outflow.
In {\refit ASP Conf. Ser. 132: Star Formation with the Infrared Space
	 Observatory}, eds. Yun, J. and Liseau, R. (ASP), pp.\ 326--329.
}

\ref{{Ceccarelli}, C., {Haas}, M. R., {Hollenbach}, D. J. and {Rudolph}, A. L. 
1997.
OI 63 Micron-determined Mass-Loss Rates in Young Stellar Objects. 
{\refit Astrophys.\ J.\/} 476:771--.
}

\ref{Cernicharo, J. and Reipurth, B. 
1996. 
Herbig-Haro Jets, CO Flows, and CO Bullets: The Case of HH 111.
{\refit Astrophys.\ J.\ Lett.\/} 460:57--61.
}

\ref{Cesaroni, R., Felli, M., Testi, L., Walmsley, C. M., and Olmi, L. 
1997.  
The disk-outflow system around the high-mass (proto)star IRAS 20126$+$4104. 
{\refit Astron.\ Astrophys.\/} 325:725--744.
}
% IRAS 20126 appears to be driving a jet seen in SiO.  

\ref{Chandler, C. J., Terebey, S., Barsony, M., Moore, T. J. T. and
Gautier, T. N. 
1996.  
Compact outflows associated with TMC-1 and TMC-1A.
{\refit Astrophys. J.\/} 471:308--320.}
% Compact outflows associated with TMC-1 and TMC-1A, gamma and outflow
% properties of these low-luminosity sources. 

\ref{Chandler, C. J., and Depree, C. G. 1995.
Vibrational Ground-State SiO J=1-0 Emission in Orion IRc2 Imaged with
the VLA. 
{\refit Astrophys.\ J.\ Lett.\/} 455:L67--L71.
}

\ref{{Chang}, C. A. and {Martin}, P. G. 
1991. 
Partially dissociative jump shocks in molecular hydrogen. 
{\refit Astrophys.\ J.\/} 378:202--213.}

\ref{{Charnley}, S. B. 
1997. 
Sulfuretted Molecules in Hot Cores.
{\refit Astrophys. J.\/} 481:396--.}

\ref{Chernin, L. M. and Wright, M. C. H. 
1996.  
High-Resolution CO observations of the molecular outflow in the Orion IRc2 region. 
{\refit Astrophys. J.\/} 467:676--683.}
% ori a irc2 collimation ratio. and flow is driven by source I. 

\ref{Chernin, L., {Masson}, C., {Gouveia Dal Pino}, E. M. and {Benz}, W. 
1994.
Momentum transfer by astrophysical jets.
{\refit Astrophys. J.\/} 426:204--214.
}

%\ref{{Chernin}, L. M. and {Masson}, C. R. 
%1995. 
%Momentum Distribution in Molecular Outflows.
%{\refit Astrophys. J.\/} 455:182--.
%}

\ref{{Chernin}, L. M. and {Masson}, C. R. 
1991. 
A nearly unipolar CO outflow from the HH 46 - 47 system.
{\refit Astrophys. J.\ Lett.\/} 382:93-96.
}

\ref{{Chieze}, J-P.,  {Pineau Des Forets}, G. and {Flower}, D. R. 
1998. 
Temporal evolution of MHD shocks in the interstellar medium. 
{\refit Mon.\ Not.\ Roy.\ Astron.\ Soc.\/} 295:672--682.}

\ref{Churchwell, E. 
1997a. 
Origin of the Mass in Massive Star Outflows.
{\refit Astrophys. J.\ Lett.\/} 479:59--61.}
% Origin of the Mass in Massive Star Outflows

\ref{Churchwell, E. 
1997b.  
Massive star formation: observational
constraints and the origin of the mass in massive outflows.
In {\refit IAU Symposium No. 182\/}, eds.\ B. Reipurth and 
C. Bertout (Kluwer Academic Publishers), pp.\ 525--536.}
% Massive star formation: observational constraints and the origin of
% the mass in massive outflows.

\ref{{Cohen}, M., {Hollenbach}, D. J., {Haas}, M. R. and {Erickson}, E. F.",
1988.
Observations of the 63 micron forbidden O I line in Herbig-Haro objects.
{\refit Astrophys.\ J.\/} 329:863--873
}

\ref{Davis, C. J., Moriarty-Schieven, G., Eisl{\"o}ffel, J., Hoare,
M. G. and Ray T. P. 
1998.  
Observations of shocked H$_2$ and 
entrained CO in outflows from luminous young stars. 
{\refit Astron. J.\/} 115:1118-1134.}
% gammas and outflow properties.  

\ref{Davis, C. J., Mundt, R., Ray, T. P., and Eisl{\"o}ffel, J. 
1995.
Near-Infrared and optical imaging of the L1551-IRS5 region -- The
importance of poorly collimated outflows from young stars.  
{\refit Astron. J.\/}110:766--775.}
% dual flow components in L1551 IRS5.  

\ref{{Davis}, C.J. and {Eisl\"offel}, J.
1995. 
Near-infrared imaging in H\_2\_ of molecular (CO) outflows from young stars.
{\refit Astron.\ Astrophys.\/} 300:851--.
}

\ref{Davis, C. J., and Smith, M. D. 1995.
Near-IR Imaging and Spectroscopy of DR21: A Case for Supersonic Turbulence.
{\refit Astron. and Astrophys. \/} 310:961--969.
}

\ref{Devine, D.,  Bally, J., Reipurth, B., Shepherd, D. S., and Watson, A. M. 
1999.  A Giant Herbig-Haro Flow from a Massive Young Star in G192.16-3.82.
{\refit Astron. J.\/} in press.}
% H_alpha, [SII], NIR Kband, H_2, [FeII] images of the G192.16 flow 
% extending 4 pc from the YSO.  

\ref{{Devine}, D., {Bally}, J., {Reipurth}, B. and {Heathcote}, S.
1997.
Kinematics and Evolution of the Giant HH34 Complex.
{\refit Astron. J.\/} 114:2095--.
}

\ref{Doyon, R. and Nadeau, D. 1988.
The Molecular Hydrogen Emission from the Cepheus A Star-Formation Region.
{\refit Astrophys.\ J.\/} 334:883--890.
}

\ref{Draine, B. T. and {McKee}, C. F. 
1993. 
Theory of interstellar shocks. 
{\refit Ann.\ Rev.\ Astron.\ Astrophys.\/} 31:373--432. }

\ref{{Dutrey}, A.,  {Guilloteau}, S. and {Bachiller}, R. 
1997. 
Successive SiO shocks along the L1448 jet axis. 
{\refit Astron.\ Astrophys.\/} 325:758--768.}

\ref{Dyson, J. E. 
1984. 
The interpretation of flows in molecular clouds.
{\refit   Astrophys.\ Space Sci.\/} 106: 181--197.
}

\ref{{Eisl\"offel}, J., {Davis}, C. J., {Ray}, T. P. and {Mundt}, R. 
1994.
Near-infrared observations of the HH 46/47 system.
{\refit Astrophys.\ J.\ Lett.\/} 422:91-93.
}

\ref{Ferreira, J.
1997.
Magnetically-driven jets from Keplerian accretion discs.
{\refit Astron.\ Astrophys.\/} 319:340--359}

\ref{Fiege, J. D. and Henriksen, R. N. 
1996a.
A global model of protostellar bipolar outflow - I.
{\refit Mon.\ Not.\ Roy.\ Astron.\ Soc.\/} 281:1038}

\ref{Fiege, J. D. and Henriksen, R. N. 
1996b.
A global model of protostellar bipolar outflow - II.
{\refit Mon.\ Not.\ Roy.\ Astron.\ Soc.\/} 281:1055}

\ref{Flower, D.R., Pineau des For\^ets, G.,
1994.
Grain-mantle erosion in magnetohydrodynamic shocks.
{\refit Mon.\ Not.\ Roy.\ Astron.\ Soc.\/} 268:724}

\ref{Flower, D.R., Pineau des For\^ets, G., Field, D., May, P.W.,
1996.
The structure of MHD shocks in molecular outflows: grain sputtering and 
SiO formation. 
{\refit Mon.\ Not.\ Roy.\ Astron.\ Soc.\/} 280:447}

\ref{{Fridlund}, C. V. M. and {Liseau}, R.
1998.
Two Jets from the Protostellar System L1551~IRS5.
{\refit Astrophys.\ J.\ Lett.\/} 599:75--.
}

\ref{Fukui, Y., Iwata, T., Mizuno, A., Bally, J., and Lane, A. P. 
1993. 
Molecular outflows. 
In {\refit Protostars \& Planets {I}{I}{I}\/}, eds.\ E. H. Levy and
J. I. Lunine (Tucson: Univ.\ of Arizona Press), pp.\ 603--639.
}
% general collimation factors for molecular outflows from low-mass
% YSOs. 

\ref{Giannakopoulou, J., Mitchell, G. F., Hasegawa, T. I., Matthews,
H. E., and Maillard, J-P. 1997.
The Star-Forming Core of Monoceros R2. 
{\refit Astrophys. J.\/} 487:346--364.
}

\ref{Garay, G., Ramirez, S. Rogr{\'\i}guez, L. F., Curiel, S.,
Torrelles, J. M. 1996.
The Nature of the Radio Sources within the Cepheus A Star-Forming
Region. 
{\refit Astrophys. J.\/} 459:193--208.
}

\ref{Garden, R. P., Hayashi, M., Gatley, I., Hasegawa, T. and Kaifu,
N. 1991.  A spectroscopic Study of the DR~21 Outflow Source. III the
CO Line Emission. 
{\refit Astrophys.\ J.\/} 374:540--554.
}

\ref{Garden, R. P., and Carlstrom, J. E. 1992.
High-Velocity HCO$^+$ Emission Associated with the DR~21 Molecular Outflow.
{\refit Astrophys.\ J.\/} 392:602--615.
}

\ref{Gayley, K.G., Owocki, S.P., Cranmer, S.R.
1995.
Momentum deposition in Wolf-Rayet winds: nonisotropic diffusion with
effective gray opacity
{\refit Astrophys.\ J.\/} 442:296--310.}

\ref{Glassgold, A. E., {Mamon}, G. A. and {Huggins}, P. J.
1989.
Molecule formation in fast neutral winds from protostars.
{\refit Astrophys.\ J.\ Lett.\/} 336:29--31.
}

\ref{Gredel, R. 
1994. 
Near-infrared spectroscopy and imaging of Herbig-Haro objects. 
{\refit  Astron.\ Astrophys.\/} 292:580--592.}

\ref{Greenhill, L. J., Gwinn, C. R., Schwartz, C., Moran, J. M., and
Diamond, P. J. 1998. 
Coexisting Conical Bipolar and Equatorial Outflows from a High-Mass
Protostar.  
{\refit Nature\/} 396:650--653.
}

\ref{{Gueth}, F., {Guilloteau}, S. and {Bachiller}, R. 
1996.
A precessing jet in the L1157 molecular outflow.
{\refit Astron.\ Astrophys.\/} 307:891--897.}

\ref{{Gueth}, F., and Guilloteau, S., 1998.
{\refit Astron.\ Astrophys.\/} in press.}

\ref{Guilloteau, S., Bachiller, R., Fuente, A. and Lucas, R. 
1992. 
First observations of young bipolar outflows with the IRAM
interferometer - 2 arcsec resolution SiO images of the molecular jet
in L 1448. 
{\refit Astron.\ Astrophys.\ Lett.\/} 265:49--52.}

\ref{Hartigan, P., Morse, J. and Raymond, J.
1994.
Mass Loss Rates, Ionization Fractions, Shock Velocities and Magnetic Fields of Stellar Jets.
{\refit Astrophys.\ J.\/} 436: 125--.
}

\ref{Hartigan, P., Carpenter, J. M., Dougados, C., and Skrutskie, M. F. 
1996.  Jet Bow Shocks and Clumpy Shells of H$_2$ Emission in the Young
Stellar Outflow Cepheus A.  
{\refit Astron.\ J.\/} 111:1278--1285.
}

\ref{Harvey, P. M., Lester, D. F., Colom{\'e}, C., Smith, B., Monin,
J-L., and Vaughlin, I. 1994.
G5.89-0.39: A Compact HII Region with a Very Dense Circumstellar Dust
Torus. 
In {\refit Astrophy. J.\/}, 433:187--198.
}

\ref{Henriksen, R.N.
1994.  
Theory of bipolar outflows.
In {\refit The Cold Universe\/}, eds.\ Th. Montmerle, Ch.J. Lada,
I.F. Mirabel, J. Tran Thanh Van (Editions Fronti\`eres: Gif-sur-Yvette),
pp.\ 241--254.}

\ref{Hollenbach, D. 
1997.
In {\refit IAU Symposium No. 182\/}, eds.\ B. Reipurth and 
C. Bertout (Kluwer Academic Publishers), pp.\ 181--198.}

\ref{Hollenbach, D. 
1985. 
Mass loss rates from protostars and OI(63 micron) shock luminosities. 
{\refit Icarus}, 61:36--39.}

\ref{Hughes, S. M. G. 1993.
Is Cepheus A East a Herbig-Haro Object? 
{\refit Astron. J.\/} 105:331--338.
}

\ref{Hunter, T. R., Phillips, T. G. and Menten, K. M. 1997.
Active Star Formation toward the Ultracompact HII Regions G45.12+0.13
and G45.07+0.13. 
{\refit Astron. J.\/} 478:283--294.
}

\ref{Kameya, O., Hasegawa, T. I., Hirano, N., Takakubo, K. 
1989.
{\refit Astrophys. J.\/} 339:222-230.}
% ngc 7538 irs1 collimation ratio.

\ref{{Kaufman}, M. J. and {Neufeld}, D.A. 
1996.
Far-Infrared Water Emission from Magnetohydrodynamic Shock Waves. 
{\refit Astrophys. J.\/} 456:611--.}

\ref{Kelly, M. L. and Macdonald, G. H. 
1996. 
Two new young stellar objects with bipolar outflows in L379. 
{\refit Mon.\ Not.\ Roy.\ Astron.\ Soc.\/} 282, 401-412.}

\ref{Lada, C. J. and Fich, M. 
1996.  
The structure and energetics of a
highly collimated bipolar outflow: NGC 2264G. 
{\refit Astrophys. J.\/} 459:638--652.}
% The structure and energetics of a highly collimated bipolar outflow:
% ngc 2264g (class 0 protostar with Lbol = 12Lsun.

\ref{{Langer}, W. D., {Castets}, A. and {Lefloch}, B. 
1996. 
The IRAS 2 and IRAS 4 Outflows and Star Formation in NGC 1333. 
{\refit Astrophys.\ J.\ Lett.\/} 471:111--115.}

\ref{{Liseau}, R.,  {Giannini}, T.,  {Nisini}, B.,  {Saraceno}, P.
       ,  {Spinoglio}, L.,  {Larsson}, B.,  {Lorenzetti}, D. and
       {Tommasi}, E.
1997.
Far-IR Spectrophotometry of HH Flows with the ISO Long-Wavelength Spectrometer. 
In {\refit IAU Symposium No. 182\/}, eds.\ B. Reipurth and 
C. Bertout (Kluwer Academic Publishers), pp.\ 111--120.
}

\ref{Mart{\'i}, J., Rodr{\'\i}guez, L. F., and Reipurth, B. 
1993. 
HH~80-81: A highly collimated Herbig-Haro complex powered by a massive
young star.  
{\refit Astrophys. J.\/} 416:208--217.}
% VLA imaging of HH80-81

\ref{Mart{\'i}, J., Rodr{\'\i}guez, L. F., and Reipurth, B. 
1995. 
Large proper motions and ejection of new condensations in the HH 80-81
thermal radio jet.  
{\refit Astrophys. J.\/} 449:184-187.}
% more VLA imaging of the central portion of the HH80-81 jet.  

\ref{Martin-Pintado, J., Bachiller, R., Fuente, A. 1992.
SiO Emission as a Tracer of Shocked Gas in Molecular Outflows.  
{\refit Astron.\ Astrophys.\/} 254:315--326.}

\ref{Masson, C. R. and Chernin, L. M. 
1992. 
Properties of swept-up molecular outflows. 
{\refit Astrophys. J.\/} 387:L47--L50.}
% power law fit to the mass spectrum

\ref{Masson, C. R. and Chernin, L. M. 
1993. 
Properties of jet-driven molecular outflows. 
{\refit Astrophys. J.\/} 414:230--241.}
% shows analytical bow-shock behavior 

\ref{Masson, C. R. and Chernin, L. M. 
1994. 
Observational constraints on outflow models.
In {\it Clouds, Cores, and Low-Mass Stars}, eds. D. Clemens and 
R. Barvainis, A.S.P.Conference Series 65:350}

\ref{McCaughrean, M.J., Rayner, J.T., Zinnecker, H., 
1994. 
Discovery of a molecular hydrogen jet near IC 348.
{\refit Ap.\ J.\/} 436:L189--L192}

\ref{McCaughrean, M. and Mac Low, M-M. 
1997.  
The OMC-1 molecular hydrogen outflow as a fragmented stellar wind bubble.  
{\refit Astron.\ J.\/} 113:391--400.}

\ref{{Mellema}, G. and {Frank}, A. 
1997.
Outflow collimation in young stellar objects. 
{\refit Mon.\ Not.\ Roy.\ Astron.\ Soc.\/} 292:795--807.
}

\ref{Menten, K. M., and Reid, M. J. 1995.
What is Powering the Orion Kleinmann-Low Nebula? 
{\refit Astrophys.\ J.\ Lett.\/} 445:L157--L160.
}

\ref{Micono, M., Davis, C.J., Ray, T.P., Eisl\"offel, J., Shetrone, M.D.
1998.
Proper motions and variability of the H$_2$ emission in the HH 46/47 system.
{\refit  Ap.\ J.\/} 494:L227--L230.}

\ref{Mitchell, G. F., Maillard, J. P., Hasegawa, T. I. 
1991.
{\refit Astrophys. J.\/} 371:342--356.}
% ngc7538 irs9 collimation ratio.

\ref{Mitchell, G. F., Lee, S. W., Maillard, J. P., Matthews, H. E., 
Hasegawa, T. I., and Harris, A. I. 
1995.  
A multitransitional CO study of GL 490. 
{\refit Astrophys. J.\/} 438:794--812.
}
% afgl 490 collimation ratio. 

\ref{{Mitchell}, G. F., {Hasegawa}, T. I., {Dent}, W. R. F., {Matthews}, H. E.
1994.
A molecular outflow driven by an optical jet.
{\refit Astrophys.\ J.\ Lett.\/} 436:177--180.
}

\ref{{Molinari}, S., {Testi}, L., {Brand}, J., {Cesaroni}, R.and {Pallo}, F. 
1998.
IRAS~23385+6053: A Prototype Massive Class 0 Object.
{\refit Astrophys. J.\ Lett.\/} 505:39--.
}

\ref{{Molinari}, S.,  {Saraceno}, P.,  {Nisini}, B.,  {Giannini}, T., 
	 {Ceccarelli}, C.,  {White}, G. J.,  {Caux}, E. and {Palla}, F.
1998.
Shocks and PDRs in an intermediate mass star forming globule: the case
of IC1396N. 
In {\refit ASP Conf. Ser. 132: Star Formation with the Infrared Space
	 Observatory}, eds. Yun, J. and Liseau, R. (ASP), pp.\ 390--.}

\ref{Moriarty-Schieven, G. H. and Snell, R. L. 
1988.  
High-resolution images of the L1551 molecular outflows. II.  Structure
and kinematics.  
{\refit Astrophys. J.\/} 332:364--378.
}

\ref{Mundt, R., Brugel, E. W. and {B\"uhrke}, T.
1987.
Jets from Young Stars: {CCD} Imaging, Long-slit Spectroscopy And Interpretation Of Existing Data.
{\refit Astrophys. J.\/} 319:275--.
}

\ref{{Nagar}, N. M., {Vogel}, S. N., {Stone}, J. M. and {Ostriker}, E. C.
1997.
Kinematics of the Molecular Sheath of the HH 111 Optical Jet.
{\refit Astrophys.\ J.\ Lett.\/} 482:195--.
}

\ref{Narayanan, G., and Walker, C. K. 1996.
Evidence for Multiple Outbursts from the Cepheus A Molecular Outflow. 
{\refit Astrophys.\ J.\/} 466:844--865.
}

\ref{{Netzer}, N., Elitzur, M.
1993.
The dynamics of stellar outflows dominated by interaction of dust and
radiation.
{\refit Astrophys.\ J.} 410:701--713.}

\ref{{Neufeld}, David A. and {Kaufman}, Michael J. 
1993. 
Radiative Cooling of Warm Molecular Gas. 
{\refit Astrophys.\ J.\/} 418:263--.}

\ref{{Nisini}, B.,  {Giannini}, T.,  {Molinari}, S.,  {Saraceno}, P.,  
	{Caux}, E.,  {Ceccarelli}, C.,  {Liseau}, R.,  {Lorenzetti}, D., 
	 {Tommasi}, E. and {White}, G. J. 
1998. 
High-J CO lines from YSOs driving molecular outflows. 
In {\refit ASP Conf. Ser. 132: Star Formation with the Infrared Space
	 Observatory}, eds. Yun, J. and Liseau, R. (ASP), pp.\ 256--.}

\ref{{Nisini}, B.,  {Lorenzetti}, D.,  {Cohen}, M.,  {Ceccarelli}, C.,  
  	{Giannini}, T.,  {Liseau}, R.,  {Molinari}, S.,  {Radicchi}, A.,  
	{Saraceno}, P.,  {Spinoglio}, L.,  {Tommasi}, E.,  {Clegg},
	P.E.,  {Ade}, P.A.R.,  {Armand}, C.,  {Barlow}, M.J.,
	{Burgdorf}, M.,  
	{Caux}, E.,  {Cerulli}, P.,  {Church}, S.E.,  {Di Giorgio},
	A.,  {Fischer}, J.,  {Furniss}, I.,  {Glencross}, W.M.,
	{Griffin}, 
	M.J.,  {Gry}, C.,  {King}, K.J.,  {Lim}, T.,  {Naylor}, 
	D.A.,  {Texier}, D.,  {Orfei}, R.,  {Nguyen-Q-Rieu},  {Sidher}, 
	S.,  {Smith}, H.A.,  {Swinyard}, B.M.,  {Trams}, N.,  {Unger}, S.J.
        and {White}, G.J. 
1996.
LWS-spectroscopy of Herbig Haro objects and molecular outflows in the
Cha II dark cloud. 
{\refit Astron.\ Astrophys.\ Lett.\/} 315:321-324.}

\ref{Padman, R., Bence, S., and Richer, J. 
1997.  
Observational properties of molecular outflows. 
In {\refit IAU Symposium No. 182\/}, eds.\ B. Reipurth and 
C. Bertout (Kluwer Academic Publishers), pp.\ 123--140.
}

\ref{{Padman}, R. and {Richer}, J. S.
1994.
Interactions between molecular outflows and optical jets.
{\refit   Astrophys.\ Space Sci.\/} 216:129--134.}

\ref{{Panagia}, N. 1991.
Ionized winds from young stellar objects. In
{\refit   The Physics of Star Formation and Early Stellar Evolution\/},
eds. C.J. Lada and N. Kylafis (Kluwer Academic Publishers), pp.\ 565--593.}

\ref{{Poetzel}, R., {Mundt}, R. and {Ray}, T. P. 
1989.
Z CMa - A large-scale high velocity bipolar outflow traced by Herbig-Haro objects and a jet.
{\refit Astron.\ Astrophys.\ Lett.\/} 224:13--16}

\ref{Raga, A.
1991.
A New Analysis of the Momentum and Mass-loss Rates of Stellar Jets.
{\refit Astron.\ J.\/} 101:1472--
}

\ref{Raga, A. C., Cant\'o, J., Calvet, N., Rodr\'{\i}guez, L. F. and
Torrelles, J.M.
1993.
A Unified Stellar Jet/Molecular Outflow Model.
{\refit Astron.\ Astrophys.\/} 276:539--.
}

\ref{{Raga}, A.C.,  {Taylor}, S.D.,  {Cabrit}, S. and {Biro}, S. 
1995.   
A simulation of an HH jet in a molecular environment. 
{\refit Astron.\ Astrophys.\/} 296:833--.}

\ref{{Reipurth}, B., {Bally}, J. and {Devine}, D.
1997.
Giant Herbig-Haro Flows.
{\refit Astron. J.\/} 114:2708--.
}

\ref{{Richer}, J.S., {Hills}, R.E. and {Padman}, R.
1992.
A fast CO jet in Orion B.
{\refit Mon.\ Not.\ Roy.\ Astron.\ Soc.\/} 254:525--538.
}

\ref{{Richer}, J.S., {Hills}, R.E., {Padman}, R. and {Russell}, A.P.G.
1989.
High-resolution molecular line observations of the core and outflow in Orion B.
{\refit Mon.\ Not.\ Roy.\ Astron.\ Soc.\/} 241: 231-246.
}

\ref{Rodr{\'\i}guez, L. F., Carral, P., Moran, J. M., and Ho, P. T. P. 
1982.
Anisotropic mass outflow in regions of star formation. 
{\refit Astrophys. J.\/} 260:635--646.}
% power law fit to the mass spectrum

\ref{Rodr{\'\i}guez, L. F., Garay, G., Curiel, S., Ram{\'\i}rez, S.,
Torrelles, J. M., G{\'o}mez, Y., and Vel{\'a}zquez, A. 1994. 
Cepheus A HW2: A powerful Thermal Radio Jet. 
{\refit Astrophys.\ J.\ Lett.\/} 430:L65--L68.
}

\ref{Rodr{\'\i}guez, L. F. 1995.
Subarcsecond Observations of Radio Continuum from Jets and Disks.  
In {\refit Revista Mexicana de Astronom{\'\i}a y Astrof{\'\i}sica,
Serie de Conferencias, Volumen 1\/}, eds.\ P. Pismis and
S. Torres-Peimbert, pp.\ 1--10.
}

\ref{Russell, A. P. G., Bally, J., Padman, R., and Hills, R. E.  
1992. 
Atomic and Molecular Outflow in DR~21. 
{\refit Astrophys. J.\/} 387:219--228.}

\ref{{Saraceno}, P., {Andr\'e}, P., {Ceccarelli}, C., {Griffin}, M.,
and {Molinari}, S. 
1996. 
An evolutionary diagram for young stellar objects. 
{\refit Astron.\ Astrophys.\/} 309:827--839.
}

\ref{{Saraceno}, P.,  {Nisini}, B.,  {Benedettini}, M., {Ceccarelli}, C.,  
	{Di Giorgio}, A. M.,  {Giannini}, T.,  {Molinari}, S., 
	 {Spinogl}, L.,  {Clegg}, P. E.,  {Correia}, J. C.,
	 {Griffin}, M. J.,  
	{Leeks}, S. J.,  {White}, G. J.,  {Caux}, E.,
	 {Lorenzetti}, D.,  
	{Tommasi}, E.,  {Liseau}, R. and {Smith}, H. A. 
1998. 
LWS Observations of Pre Main Sequence Objects. 
In {\refit ASP Conf. Ser. 132: Star Formation with the Infrared Space
	 Observatory}, eds. Yun, J. and Liseau, R. (ASP), pp.\ 233--.
}

\ref{Schulz, A., Henkel, C., Beckmann, U., Kasemann, C., Schneider, G.
Nyman, L. A., Persson, G., Gunnarsson, G. G., and Delgado, G. 1995. 
A High-Resolution CO J=4-3 Map of Orion-KL. 
{\refit Astron. and Astrophys.\/} 295:183--193.
}

\ref{Shepherd, D. S. and Churchwell, E. 
1996. 
Bipolar molecular outflows in massive star-formation regions. 
{\refit Astrophys. J.\/} 472:225--239.}
% single-dish maps of massive bipolar outflows and Lbol vs Mdot

\ref{Shepherd, D. S., Watson, A. M., Sargent, A. I. and Churchwell, E. 
1998.
Outflows and luminous YSOs: A new perspective on the G192.16 massive bipolar outflow.  
{\refit Astrophys. J.\/} accepted.}
% The molecular outflow and IR images of G192.16

\ref{Shepherd, D. S., Kurtz, S. E. 
1998. 
{\refit In preparation.\/}}
% The disk and YSO in G192.16

\ref{{Shu}, F. H. and {Shang}, H. 
1997.
Protostellar X-rays, Jets, and Bipolar Outflows. 
In {\refit IAU Symposium No. 182\/}, eds.\ B. Reipurth and 
C. Bertout (Kluwer Academic Publishers), pp.\ 225--239.}

\ref{{Shu}, F.H., {Ruden}, S.P., {Lada}, C.J., and {Lizano}, A.
Star formation and the nature of bipolar outflows.
1991.
{\refit Astrophys. J. L.\/} 370:31-34.}

\ref{{Shu}, F., {Najita}, J., {Ostriker}, E.,{Wilkin}, F., {Ruden}, S. and {Lizano}, S.
1994.
Magnetocentrifugally driven flows from young stars and disks. 1: A generalized model.
{\refit Astrophys. J.\ Lett.\/} 429:781-796.
}

\ref{{Smith}, M.D. 
1994.
Jump shocks in molecular clouds - speed limits and excitation levels. 
{\refit Mon.\ Not.\ Roy.\ Astron.\ Soc.\/} 266:238--.}

\ref{Smith, M. D., Suttner, G. and Yorke, H. W. 
1997.  
Numerical hydrodynamic simulation of jet-driven bipolar outflows. 
{\refit Astron.\ Astrophys.\/} 323:223--230.
}
% Numerical hydrodynamic simulation of jet-driven bipolar outflows.
% Simulating the ngc2264g and cephE flows, gamma vs time and i. 

\ref{{Smith}, M. D., {Eisl\"offel}, J. and {Davis}, C. J. 
1998.
ISO observations of molecular hydrogen in the DR21 bipolar outflow. 
{\refit Mon.\ Not.\ Roy.\ Astron.\ Soc.\/} 297:687--691.}

\ref{{Suttner}, G.,  {Smith}, M.D.,  {Yorke}, H.W. and {Zinnecker}, H.
1997.
Multi-dimensional numerical simulations of molecular jets. 
{\refit Astron.\ Astrophys.\/} 318:595--607.}

\ref{Stahler, S. W. 
1994.  
The kinematics of molecular outflows.
{\refit Astrophys. J.\/} 422:616--620.}
% power law fit to the mass spectrum and turbulent entrainment in jets
% theory. 

\ref{{Stone}, J. M., {Xu}, J. and {Mundy}, L.G.
1995. 
Formation of bullets by hydrodynamical instabilities in stellar outflow. 
{\refit Nature\/} 377:315--.
}

\ref{{Tafalla}, M. and {Bachiller}, R.
1995. 
Ammonia emission from bow shocks in the L1157 outflow. 
{\refit Astrophys. J.\ Lett.\/} 443:37-40.}

\ref{{Taylor}, S. D. and {Raga}, A. C.
1995.
Molecular mixing layers in stellar outflows.
{\refit Astron.\ Astrophys.\/} 296:823--.
}

\ref{Torrelles, J. M., Verdes-Montenegro, L., Ho, P. T. P.
Rodr{\'\i}guez, L. F., and Jorge, C. 1993.
>From Bipolar to Quadrupolar -- The Collimation Processes of the
Cepheus A Outflow.  
{\refit Astrophys.\ J.\/} 410:202--217.
}

\ref{van Dishoeck, E. F., {Blake}, G. A., {Jansen}, D. J. and {Groesbeck}, T. D.",
1995. 
Molecular Abundances and Low-Mass Star Formation. II. Organic and
Deuterated Species toward IRAS 16293-2422. 
{\refit Astrophys.\ J.\/} 447:760--.}

\ref{van Dishoeck E.F. and Blake G.A. 
1998.
{\refit Ann.\ Rev.\ Astron.\ Astrophys.\/} in press.}

\ref{Velusamy, T. and Langer, W. D. 
1998. 
Outflow-Infall interactions as a mechanism for terminating accretion in protostars.  
{\refit Nature\/} 392:685-687.
}

\ref{V\"olker, R., Smith, M.D., Suttner, G., Yorke, H.W. 
1999.
Numerical hydrodynamical simulations of molecular outflows driven
by hammer jets.
{\refit  Astron.\ Astrophys.\ }, in press}

\ref{Wilner, D. J., Welch, W. J., Forster, J. R. 1992.
The G5.89-0.39 UC HII Region: Kinematics and Millimeter Aperture
Synthesis Maps. 
{\refit Am.\ Astron.\ Soc.\ Meeting, 181\/}, number 22.06.
}

\ref{Wright, M. C. H., Plambeck, R. L., and Wilner, D. J. 1996. 
A Multiline Aperture Synthesis Study of Orion-KL. 
{\refit Astrophys.\ J.\/} 469:216--237.
}

\ref{Wolf, G. A., Lada, C. J., and Bally, J. 1990.
The Giant Molecular Outflow in Mon R2.
{\refit Astron. J.\/}, 100:1892--1902.
}

\ref{Wood, D. O. S. 1993.
Bipolar Molecular Outflow from a Massive Star: High Resolution
Observations of G5.89-0.39. 
In {\refit ASP conf. Series\/}, 35:108--110.
}

\ref{{Wright}, C.M.,  {Drapatz}, S.,  {Timmermann}, R.,  {Van Der
	Werf}, P.P., {Katterloher}, R. and {De Graauw}, T.
1996. 
Molecular hydrogen observations of Cepheus A West.
{\refit Astron.\ Astrophys.\ Lett.\/} 315:301--L304.}

\ref{{Wu}, Y., {Huang}, M. and {He}, J.
1996.
A catalogue of high velocity molecular outflows.
{\refit Astron.\ Astrophys.\ Suppl.\/} 115:283--.
}

\ref{Yamashita, T., Suzuki, Kaifu, N., Tamura, M., Mountain,
C. M. and Moore, T. J. T. 
1989.  
A new CO bipolar flow and dense disk system 
associated with the infrared reflection nebula GGD 27 IRS.
{\refit Astrophys. J.\/} 373:560--566.}
% the co flow in HH80-81.

\ref{{Zhang}, Q., {Hunter}, T. R. and {Sridharan}, T. K.
1998.
A Rotating Disk around a High-Mass Young Star.
{\refit Astrophys.\ J.\ Lett.\/} 505:151-154.
}    

\ref{Zijlstra, A. A., Pottasch, S. R., Engels, D., Roelfsema, P. R.,
Hekkert, P. L., and Umana, G. 1990. 
Mapping the Outflow of OH5.89-0.39. 
{\refit Mon.\ Not.\ Roy.\ Astron.\ Soc.\/} 246:217--236.
}

\end